\documentclass[5p,times,twocolumn]{elsarticle}

\usepackage{amssymb}
\usepackage{amsmath}
\usepackage{amsfonts}
\usepackage{graphicx}
\usepackage{color}
\usepackage{tabularx}
\usepackage{bm}
\usepackage{booktabs}
\usepackage{textcomp}
\usepackage{xcolor}
\usepackage{hyperref}
\usepackage{multirow}
\usepackage{url}
\usepackage{algorithm}
\usepackage{algpseudocode}
\usepackage{placeins}
\usepackage{float}
\usepackage{dblfloatfix}
\usepackage{natbib}
\usepackage{titlesec}
\usepackage{enumitem}
\hypersetup{hidelinks=true}

\titleformat{\subsection}
{\bfseries\itshape}
{\thesubsection}
{1em}
{}

\journal{Biomedical Signal Processing and Control}

\begin{document}

\begin{frontmatter}

\title{Sparse-View Lung Nodule Volumetry from Digitally Reconstructed Radiographs via AReT: Anatomy-Regularized TensoRF}

\author{Spoorthi M}
\author{Suja Palaniswamy\corref{cor1}}
\ead{p\_suja@blr.amrita.edu}
\cortext[cor1]{Corresponding author: Suja Palaniswamy (email: p\_suja@blr.amrita.edu).}

\affiliation{organization={Department of Computer Science \& Engineering, Amrita School of Computing, Bengaluru, Amrita Vishwa Vidyapeetham},
             country={India}}

\begin{abstract}
We identify and resolve a previously unreported failure mode in TensoRF when applied to X-ray attenuation fields: the default density shift of $-10$, originally introduced for RGB scene reconstruction, suppresses density gradients and prevents sparse-view medical reconstruction regardless of learning rate or regularization strategy. Setting the density shift to zero restores gradient flow and enables stable volumetric reconstruction of pulmonary nodules from only three orthogonal X-ray projections.

Building on this observation, we propose AReT, an anatomy-regularized tensorial radiance field framework for three-dimensional lung nodule reconstruction using coronal, sagittal, and axial projections generated from the LIDC-IDRI dataset containing 19 patients with radiologist-annotated nodules. Unlike existing neural radiance field approaches that require dense multi-view acquisition or inadequately model attenuation physics, the proposed framework is specifically designed for sparse-view thoracic imaging. To enforce physiologically plausible attenuation distributions, we incorporate chest-anatomy-aware regularization combining $\ell_1$ sparsity and total variation smoothness, leveraging the predominantly air-filled structure of the thoracic cavity.

A systematic comparison across 11 reconstruction strategies demonstrates that anatomy-aware regularization consistently outperformed the generative-prior-guided approaches evaluated here evaluated here. Evaluated against radiologist consensus segmentations, the proposed method achieved a Pearson correlation of $r = 0.983$ ($p < 0.0001$) for clinically actionable nodules $\geq 10$,mm ($n = 14$), with a median absolute volumetric error of 11.4\%, near-zero systematic bias of $-77.3$,mm$^3$, and an 8.4$\times$ improvement over spherical volume approximation. Threshold selection follows a fixed percentile sweep applied uniformly across all patients, while final candidate selection uses annotated nodule diameter as a semi-assisted post-processing size filter.
\end{abstract}

\begin{keyword}
pulmonary nodule volumetry \sep
neural radiance fields \sep
sparse-view reconstruction \sep
TensoRF \sep
digitally reconstructed radiographs \sep
anatomy-aware regularisation \sep healthcare AI
\end{keyword}

\end{frontmatter}

\section{Introduction}
\label{sec:introduction}

Lung cancer causes $\sim$1.8M deaths annually~\cite{sung2021global}. Accurate pulmonary nodule volumetry is critical for Tumor, Node, Metastasis (TNM) staging, radiotherapy planning, and longitudinal monitoring under Lung Imaging Reporting and Data System (Lung-RADS) and Fleischner guidelines~\cite{lung_rads2019,macmahon2017guidelines}. Volume is inherently more sensitive than diameter to interval change (a 20\% diameter change corresponds to a 73\% volume change~\cite{revel2004diameters}), yet accurate 3D volumetry requires dense CT acquisitions with high radiation burden. Thus there is a pressing need for accurate 3D nodule reconstruction from minimal projections.

Classical sparse-view methods (Feldkamp–Davis–Kress (FDK), Simultaneous Algebraic Reconstruction Technique (SART)) require 20–50 views and are unlikely to produce clinically meaningful reconstructions at ultra-sparse regimes such as three projections~\cite{feldkamp1984, andersen1984}. Neural Radiance Fields (NeRF)~\cite{mildenhall2020nerf} have inspired medical adaptations NAF~\cite{Zha2022} (requiring 50--100 Cone Beam Computed Tomography (CBCT) projections) and MedNeRF~\cite{abril2022mednerf} (population-level, not patient-specific) but no method has demonstrated volumetrically accurate nodule reconstruction from as few as three orthogonal X-rays, and Tensorial Radiance Fields (TensoRF) remain unstudied in medical attenuation fields~\cite{chen2022tensorf}.

We propose AReT, an anatomy-regularized TensoRF framework for 3D lung nodule volumetry from three orthogonal digitally reconstructed radiographs (DRRs: coronal, sagittal, axial), evaluated on LIDC-IDRI (1{,}018 thoracic CTs, up to four radiologist annotations per scan)~\cite{armato2011lidc}. We identify and resolve a critical TensoRF failure mode: the default density shift ($\Delta=-10$) suppresses all learned density; setting $\Delta=0$ restores gradient flow. We further introduce a thorax-specific $\ell_1$+TV regularisation scheme and a post-hoc density-gradient uncertainty volume for boundary confidence estimation without additional training.

It is important to clarify a paradigm distinction that bears on the interpretation of our evaluation. 
The proposed TensoRF framework is a \textit{scene-specific} (equivalently, patient-specific) implicit 
neural representation: each model is trained from random initialisation on three DRRs belonging to a 
single patient, with no cross-patient weight sharing, no population-level latent code, and no joint 
training across cases. This is architecturally analogous to iterative CT reconstruction algorithms 
such as FDK or SART, where each reconstruction is an 
independent inverse problem. Consequently, the $n = 19$ evaluation cohort characterises the 
\textit{breadth of independent test cases} rather than a training sample size, and classical train/test overfitting does not apply in the standard sense, because optimisation is scene-specific rather than population-trained. However, analogous risks remain including fitting to reconstruction artefacts, exploiting simulator-specific statistics, or over-specialisation to noiseless DRRs and should be acknowledged as limitations of the current evaluation.

This work is explicitly framed as a proof-of-concept study on a limited cohort of 19 patients; all conclusions regarding reconstruction behaviour, robustness, and methodological performance are strictly scoped to this evaluation and not generalised beyond it without validation on a larger and more morphologically diverse cohort.

A systematic 11-strategy evaluation on LIDC-IDRI is demonstrated in Section~\ref{sec:experiments}, showing that anatomy-regularised reconstruction outperforms the generative-prior-guided approaches evaluated here for patient-specific sparse-view volumetry. The contributions of this work are categorised and briefed as follows:
 
\textbf{Primary methodological novelty}
\begin{enumerate}
  \item \textbf{Critical density parameterisation fix.}
    The default $\Delta = -10$ yields zero density throughout
    training; $\Delta = 0$ is necessary, and empirically sufficient in this setting, for
    sparse-view medical reconstruction.
    This failure mode does not appear to have been previously documented in the medical imaging TensoRF literature and transfers
    directly to any TensoRF deployment in medical attenuation fields.
\end{enumerate}

\textbf{Integration / adaptation contributions}
\begin{enumerate}
  \setcounter{enumi}{1}
  \item \textbf{Anatomy-informed regularisation.}
    $\ell_1$ sparsity (exploiting air-dominant thoracic composition)
    combined with total variation smoothness enables stable 3D
    reconstruction from only three DRRs.
    Although $\ell_1$+TV regularization is well-established in
    sparse-signal recovery, its deployment as a thorax-specific
    inductive bias within the corrected TensoRF framework, and its
    empirical superiority over generative-prior alternatives in this
    setting, constitute the integration contribution.
  \item \textbf{Systematic 11-strategy evaluation} on LIDC-IDRI, showing
    anatomy-regularized reconstruction outperforms the
    generative prior-guided approaches evaluated here for
    patient-specific sparse-view volumetry.
 
\end{enumerate}

On 19 LIDC-IDRI patients, the method achieves Pearson $r=0.983$ ($p<0.0001$) on clinically actionable nodules $\geq10$\,mm ($n=14$), median absolute volumetric error 11.4\%, near-zero bias ($-43.6$\,mm$^3$, Bland-Altman), and an $8.4\times$ improvement over spherical approximation.

It is further noted that, consistent with prior sparse-view medical 
NeRF works, training and evaluation are 
conducted on digitally reconstructed radiographs (DRRs) simulated 
from CT via the Beer--Lambert forward model, rather than real 
acquired chest X-rays~\cite{Corona2022}. This choice is necessitated by the 
requirement for paired projection and volumetric ground-truth data: 
the LIDC-IDRI dataset provides annotated CT volumes but no 
corresponding clinical radiographs. While DRR simulation faithfully 
models primary X-ray attenuation, real radiographs additionally 
exhibit scatter, detector noise, beam hardening, motion artefacts, 
and acquisition inconsistencies not present in simulation. The 
domain gap between simulated and real projections represents a 
known challenge for clinical translation, which we explicitly 
acknowledge and identify as a primary direction for future work.

The paper is organised as follows: Section~\ref{sec:related} elaborates literature survey, Section~\ref{sec:dataset} describes the dataset, Section~\ref{sec:methodology} explains the proposed framework, Section~\ref{sec:experiments} narrates the experimental setup, Section~\ref{sec:results} illustrates results and Section~\ref{sec:discussion} provides discussion followed by conclusion in Section~\ref{sec:conclusion}.

\section{Literature Survey}
\label{sec:related}
 
\subsection*{\textbf{Classical Sparse-View and Compressed-Sensing CT}}
FDK and SART are the
clinical standards for CT reconstruction but require 20--50 views;
both produce severe streak artefacts below this threshold.
Statistical iterative reconstruction (SIR) methods incorporating TV
or Huber regularisation reduce
artefacts at moderate undersampling (10--30 views) and underpin
commercial low-dose CT pipelines~\cite{Rudin1992,Sidky2008}.
Compressed sensing (CS) provides
theoretical guarantees for recovery of sparse signals from
underdetermined measurements when the restricted isometry
property (RIP) holds~\cite{Candes2006,Donoho2006}; TV-minimisation over the image domain
enables accurate reconstruction from 20--60 projections on
thoracic data~\cite{Chen2008SB}.
However, three orthogonal line-integral operators do not satisfy
the RIP at clinically relevant resolutions, making CS-CT unlikely to yield diagnostically meaningful reconstructions at the three-projection regime studied here, regardless of regularisation strength.
The proposed $\ell_1$+TV penalties share conceptual heritage
with CS-CT but are applied within a neural implicit representation
that parameterises the density field more flexibly than a fixed
voxel grid.

\subsection*{\textbf{Physics-Informed Reconstruction}}
The Beer--Lambert model has served as a differentiable forward
operator since early emission tomography~\cite{Shepp1982}.
Physics-informed neural networks (PINNs) and
learned primal-dual frameworks achieve
state-of-the-art sparse-view CT results by embedding the
projection operator as a physics constraint~\cite{Raissi2019, Adler2018}.
The proposed method shares this spirit: Beer--Lambert governs the
photometric loss, and the anatomy-aware regularisation encodes
thorax-specific priors.
The key distinction is that no cross-patient training data is
required; the physics constraint and anatomical prior are
sufficient for patient-specific reconstruction from three views.
\vspace{-0.1cm}
\subsection*{\textbf{Neural Radiance Fields}}
NeRF maps 3D coordinates to colour and
density via an MLP with differentiable volume rendering, but
requires dense multi-view input.
TensoRF replaces the MLP with vector-matrix
(VM) tensorial decomposition, substantially reducing memory
and training time.
The density parameterisation failure identified here (the default
$\Delta=-10$ suppressing all gradients in the X-ray attenuation
setting) is specific to TensoRF's initialisation scheme and does
not affect MLP-based NeRFs, which initialise density near zero
by default.
Sparse-view NeRF methods address the multi-view requirement
through explicit regularisation: RegNeRF
regularizes unseen viewpoints and depth maps from as few as
three RGB views~\cite{Niemeyer2022}; DietNeRF adds semantic
consistency~\cite{Jain2021}.
These methods operate in the RGB domain and are not directly
applicable to X-ray attenuation, but the principle that explicit
constraints on unobserved views are necessary for sparse-view
NeRF is central to the proposed AReT framework.
 
\subsection*{\textbf{Medical NeRF Adaptations}}
NAF adapts NeRF to X-ray attenuation for
sparse-view CBCT, requiring 50--100 projections.
MedNeRF embeds NeRF within a GAN
framework (Generative Radiance Fields (GRAF)) for single-X-ray
volumetric estimation, but the latent code is sampled from a
population-level prior rather than inverted from patient
observations, producing diffuse non-calibrated density fields~\cite{schwarz2020graf}.
To our knowledge, no prior published method has demonstrated has demonstrated volumetrically accurate nodule reconstruction from as few as three orthogonal X-rays, and this is among the first works to apply TensoRF to medical attenuation fields.
 
\subsection*{\textbf{Generative and Diffusion-Based Reconstruction}}
Score-based diffusion models applied to sparse-view
CT demonstrate impressive image quality at
moderate undersampling~\cite{Chung2022}.
These methods share a characteristic limitation for patient-specific
volumetry: the generative prior encodes average anatomical
statistics; latent inversion at inference is weakly constrained by
sparse observations, yielding population-average rather than
individual morphology.
This limitation is consistent with the ablation result in
Section~\ref{sec:res_tensorf}: MedNeRF prior injection consistently
degrades accuracy relative to the prior-free patient-specific
approach.
 
\subsection*{\textbf{Gaps Addressed}}
Three gaps have motivated this work:
(i)~\textit{Dense view assumption}---existing NeRF methods require
tens to hundreds of views;
(ii)~\textit{Population-level priors}---generative approaches yield
average statistics, not patient-specific morphology;
(iii)~\textit{Absent physiological constraints}---without
thorax-specific priors, networks distribute density diffusely,
preventing focal lesion isolation.
The proposed framework addresses all three.
The 11 evaluated strategies are a systematic diagnostic
ablation, not competitive external baselines; each isolates one
design decision in the progression to the proposed method.
Classical methods (FDK, SART, CS-CT) are designed for
20--100 projections and have not been validated at three
views; NAF requires 50--100 projections; MedNeRF uses
population priors.
Direct numerical comparison with methods requiring
different input regimes is therefore not straightforward.   
\section{Dataset}
\label{sec:dataset}

Experiments are conducted on the Lung Image Database Consortium and
Image Database Resource Initiative (LIDC-IDRI)
dataset, the largest publicly available
repository of annotated pulmonary CT acquisitions~\cite{armato2011lidc}. The dataset
comprises of $1{,}018$ thoracic CT scans with nodule annotations
provided by up to four independent radiologists per scan.

\begin{figure}[!t]
    \centering
    \includegraphics[width=1.0\linewidth]{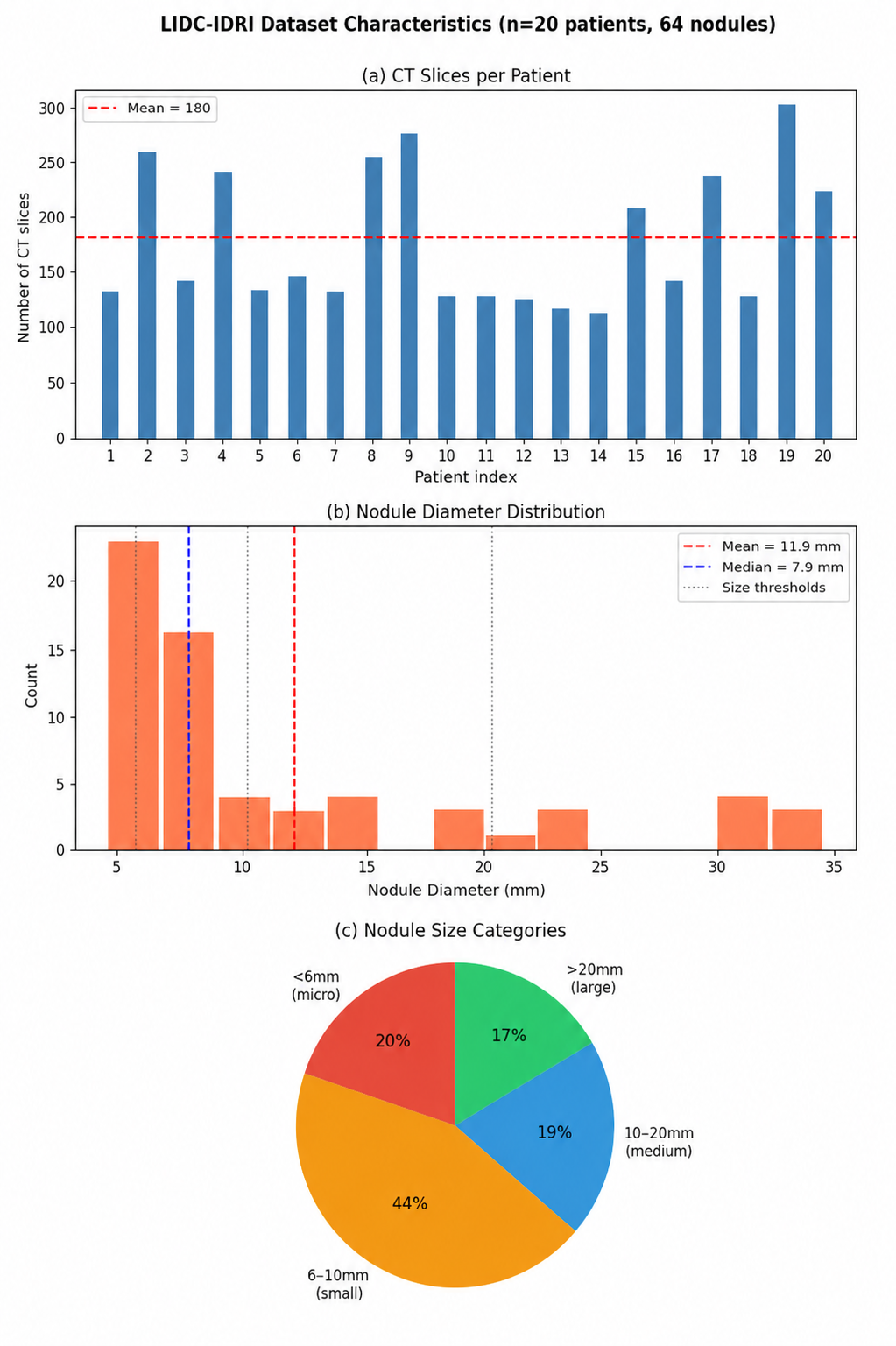}
    \caption{Dataset statistics for the LIDC-IDRI subset ($n=19$
    patients). Nodules span a wide range of sizes and volumes,
    stratified into small ($<10$\,mm), medium (10--20\,mm), and
    large ($>20$\,mm) categories.}
    \label{fig:dataset_stats}
\end{figure}

For this study, a subset of 19 patients was selected, spanning a
clinically representative range of nodule sizes. All scans were
acquired on GE Medical Systems scanners (in-plane pixel spacing
$0.729 \pm 0.084$\,mm, slice thickness $2.00 \pm 0.63$\,mm).
Nodules are stratified into three size categories following
Fleischner Society guidelines:
small ($<10$\,mm, $n=5$), medium ($10$--$20$\,mm, $n=6$), and
large ($>20$\,mm, $n=8$), as summarised in
Table~\ref{tab:patients}. For patients with multiple annotated
nodules, the largest nodule by consensus volume is selected as
the primary evaluation target, consistent with clinical
prioritisation of the dominant lesion. Dataset statistics are
shown in Fig.~\ref{fig:dataset_stats}.

\begin{table}[ht]
\centering
\caption{Patient cohort summary. GT volume and diameter were derived
from \texttt{pylidc} consensus masks at $\text{clevel}=0.5$.}
\label{tab:patients}

\footnotesize
\renewcommand{\arraystretch}{1.15}
\setlength{\tabcolsep}{5pt}

\begin{tabular*}{\linewidth}{@{\extracolsep{\fill}}ccccc@{}}
\toprule
\textbf{ID} & \textbf{Diam.} &
\textbf{Vol.} & \textbf{Ann.} &
\textbf{Cat.} \\
 & (mm) & (mm$^3$) &  &  \\
\midrule
0001 & 32.8 & 6708.8 & 4 & L \\
0002 & 30.8 & 8277.5 & 2 & L \\
0003 & 31.0 & 5452.3 & 4 & L \\
0004 &  6.8 &   61.7 & 4 & S \\
0005 &  7.9 &  153.2 & 4 & S \\
0006 & 15.1 &  379.9 & 4 & M \\
0007 & 34.4 & 7797.2 & 4 & L \\
0008 &  8.4 &  112.9 & 4 & S \\
0009 &  7.4 &   58.0 & 1 & S \\
0010 & 10.0 &  167.0 & 4 & M \\
0011 & 15.3 &  723.0 & 4 & M \\
0012 & 12.7 &  820.8 & 4 & M \\
0013 & 22.0 & 1783.6 & 4 & L \\
0014 & 17.8 & 1479.5 & 4 & M \\
0015 & 23.5 & 3189.2 & 4 & L \\
0016 & 18.6 & 1519.2 & 4 & M \\
0017 &  7.8 &  118.8 & 4 & S \\
0018 & 21.8 & 2841.0 & 4 & L \\
0019 & 31.7 & 7198.8 & 2 & L \\
\midrule
\multicolumn{5}{@{}l@{}}{\textit{S}: $n=5$, median $7.8$\,mm} \\
\multicolumn{5}{@{}l@{}}{\textit{M}: $n=6$, median $16.5$\,mm} \\
\multicolumn{5}{@{}l@{}}{\textit{L}: $n=8$, median $31.3$\,mm} \\
\bottomrule
\end{tabular*}

\end{table}

\section{Methodology}
\label{sec:methodology}

\subsection{Overview}
\label{sec:overview}

\begin{figure}[htbp]
    \centering
    \includegraphics[width=\linewidth]{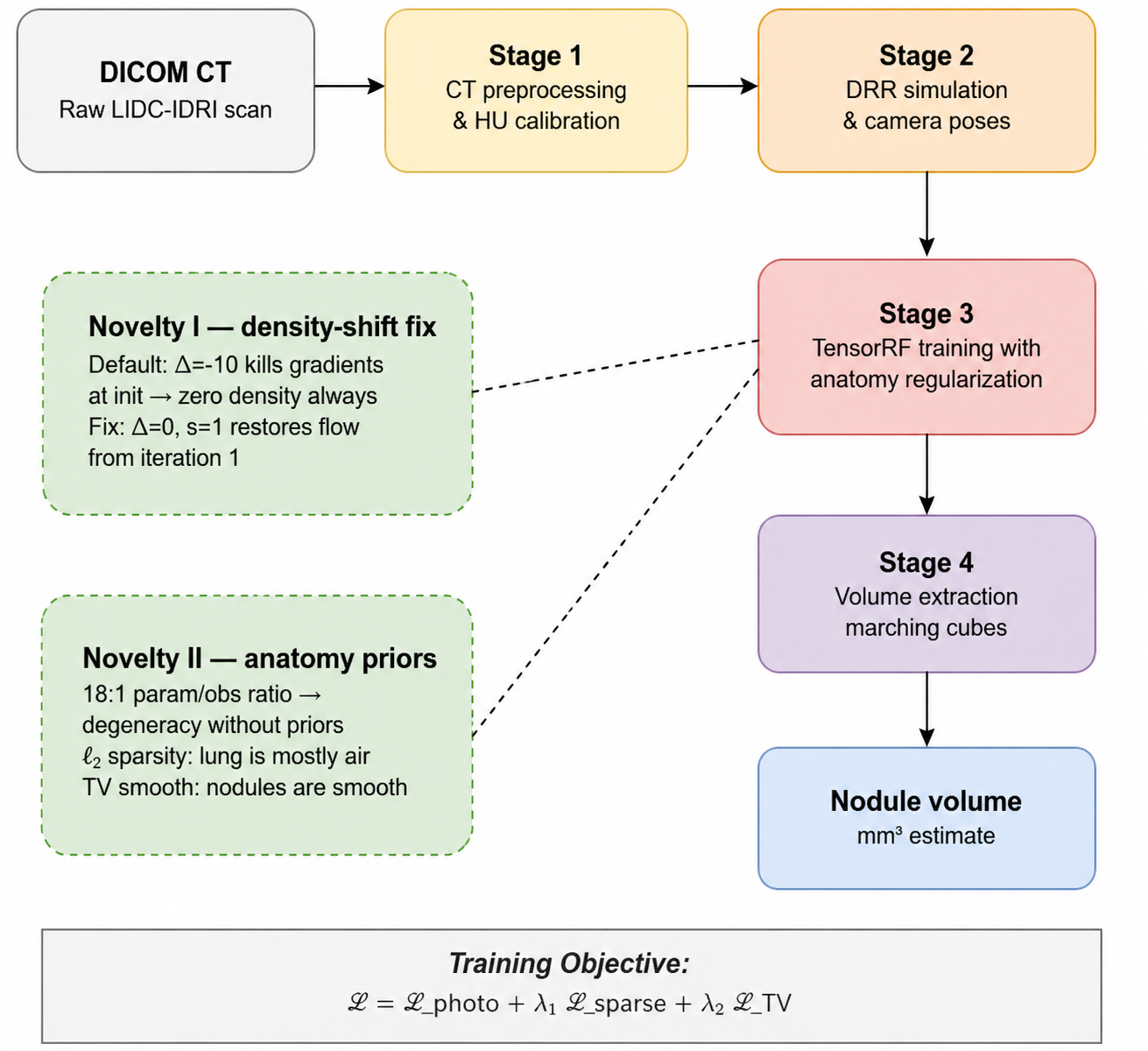}
    \caption{Overview of the proposed AReT NeRF framework for sparse-view pulmonary
    nodule volumetry. The pipeline consists of:
    (1) CT preprocessing and Hounsfield-unit calibration;
    (2) Beer--Lambert DRR simulation and assignment of
    coronal, sagittal, and axial camera poses;
    (3) anatomy-regularized TensoRF training with the
    corrected density parameterization; and
    (4) volumetric reconstruction and nodule extraction
    using marching cubes and connected-component analysis.}
    \label{fig:pipeline}
\end{figure}

We propose AReT, an anatomy-regularized tensorial NeRF framework for 3D lung nodule volumetry from three orthogonal digitally reconstructed radiographs (DRRs: coronal, sagittal, axial). The pipeline proceeds through four stages (Fig.~\ref{fig:pipeline}): (i) CT preprocessing and HU calibration; (ii) DRR simulation and camera pose assignment via Beer-Lambert forward projection; (iii) anatomy-regularized TensoRF training with corrected density parameterization; and (iv) volume extraction via marching cubes and connected component analysis. The two principal novelties are the identification and resolution of a density-shift initialization failure (Section~\ref{sec:densityshift}) and an anatomy-aware regularization strategy (Section~\ref{sec:regularization}).

\subsection{CT Data Preprocessing}
\label{sec:preprocessing}

\begin{figure*}[ht]
    \centering
    \includegraphics[width=\linewidth]{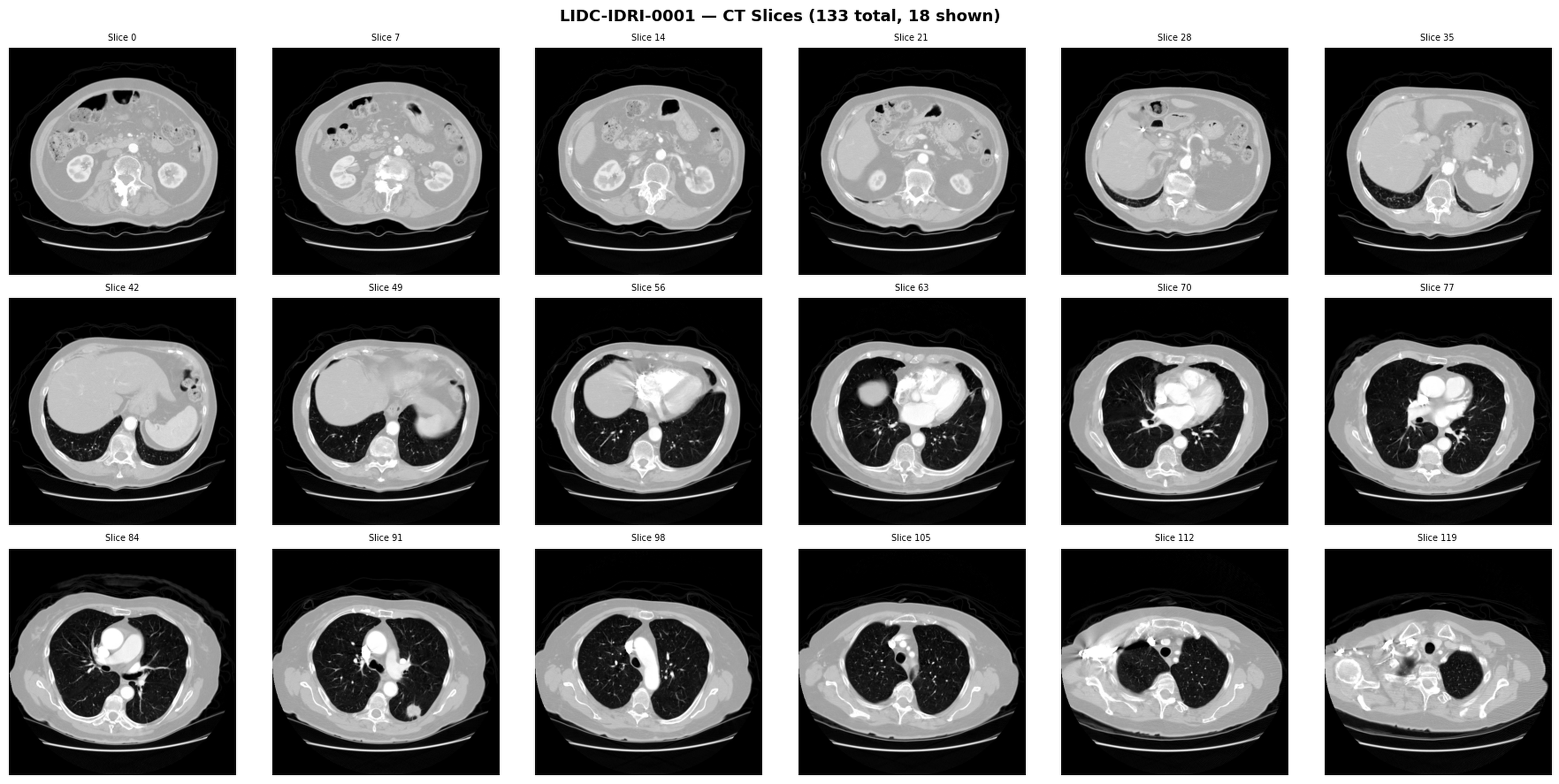}
    \caption{Representative axial CT slices from the preprocessed
    LIDC-IDRI volume (\texttt{LIDC-IDRI-0001}). Slices are shown after
    conversion to Hounsfield Units, lung-window clipping to
    $[-1000, 400]$\,HU, and linear normalisation to $[0,1]$. The
    preprocessing pipeline preserves pulmonary anatomy and soft-tissue
    contrast while suppressing high-density structures outside the
    diagnostic lung window.}
    \label{fig:ct_preprocessing}
\end{figure*}

DICOM slices are loaded via \texttt{pydicom}, sorted by \texttt{ImagePositionPatient} Z-coordinate, and stacked into a $(Z,H,W)$ array. Pixel values are converted to Hounsfield Units:
\begin{equation}
    \text{HU} = p \cdot s + b
    \label{eq:hu}
\end{equation}
where $p$ is the raw pixel value, $s$ is \texttt{RescaleSlope}, and $b$ is \texttt{RescaleIntercept}. HU values are clipped to $[-1000, 400]$ (air to soft tissue) and linearly normalised to $[0,1]$, suppressing bone artefacts while preserving soft-tissue contrast. To verify volumetric consistency after preprocessing, evenly spaced
axial slices were visually inspected across the reconstructed CT
volume. Fig.~\ref{fig:ct_preprocessing} illustrates representative
slices after Hounsfield Unit conversion, lung-window clipping, and
linear normalisation, demonstrating preservation of anatomical
structures and soft-tissue contrast throughout the scan.

\subsection{DRR Simulation and Camera Pose Assignment}
\label{sec:drr}

\begin{figure}[!htb]
    \centering
    \includegraphics[width=\linewidth]{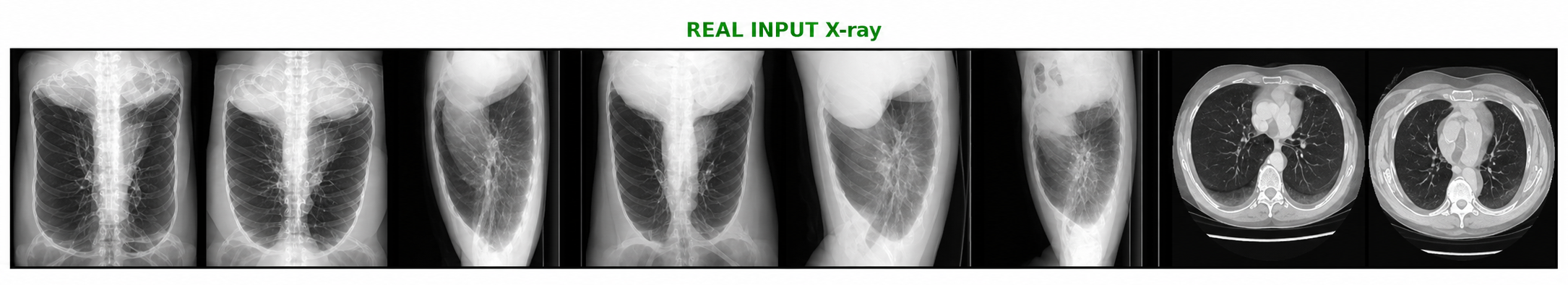}
    \caption{Representative DRRs from LIDC-IDRI-0001}
    \label{fig:drrs}
\end{figure}

DRRs are synthesised via the Beer-Lambert law:
\begin{equation}
    I(\mathbf{r}) = I_0 \cdot \exp\!\left(-\int \mu(\mathbf{x})\,\mathrm{d}x\right)
    \label{eq:beer}
\end{equation}
where $\mu(\mathbf{x})$ is the linear attenuation coefficient, approximated by summing normalised HU values along each projection direction. Three orthogonal projections are rendered at $128\times128$ pixels (Fig.~\ref{fig:drrs}), with camera-to-world transforms:

\begin{equation}
\setlength{\arraycolsep}{2pt}
\small
\begin{aligned}
\mathbf{C}_{\mathrm{cor}} &=
\begin{pmatrix}
1&0&0&0\\
0&0&1&d\\
0&-1&0&0\\
0&0&0&1
\end{pmatrix}
\hspace{2mm}
\mathbf{C}_{\mathrm{sag}} =
\begin{pmatrix}
0&0&-1&d\\
1&0&0&0\\
0&-1&0&0\\
0&0&0&1
\end{pmatrix}
\\[2mm]
\mathbf{C}_{\mathrm{ax}} &=
\begin{pmatrix}
1&0&0&0\\
0&1&0&0\\
0&0&1&d\\
0&0&0&1
\end{pmatrix}
\end{aligned}
\label{eq:poses}
\end{equation}
where $d=50.0$ is the camera distance and focal length $f=1280$ pixels ($\approx10^\circ$ FoV). Ray directions for pixel $(i,j)$ are:
\begin{equation}
    \mathbf{d}_{ij} = \mathrm{normalize}\!\left(
    \frac{i - W/2}{f},\;
    -\frac{j - H/2}{f},\;
    -1
\right)
    \label{eq:raydirs}
\end{equation}
yielding $49{,}152$ training rays per patient ($3\times128\times128$).

\subsubsection*{\textbf{Simulation fidelity and domain gap.}}
The DRR forward model (Eq.~\ref{eq:beer}) approximates primary 
X-ray attenuation under monoenergetic, scatter-free, motion-free 
conditions. Real clinical chest radiographs deviate from this 
idealisation in several respects: (i)~\textit{scatter radiation}, 
which adds a spatially varying background signal not captured by 
line-integral projection; (ii)~\textit{beam hardening}, arising 
from the polyenergetic X-ray spectrum interacting with 
heterogeneous tissue compositions; (iii)~\textit{detector noise 
and response}, including quantum noise at clinical dose levels and 
detector point-spread functions; (iv)~\textit{motion artefacts} 
from cardiac pulsation and respiratory motion during acquisition; 
and (v)~\textit{anatomical overlap}, which is present in both 
simulated and real projections but may manifest differently under 
real acquisition geometry. The present work operates entirely 
within the DRR domain and does not claim robustness to these 
real-world degradations. Domain adaptation from simulated to real projections is identified as a critical open problem for clinical translation. The gap between idealised DRR conditions and real radiographic deployment is substantial, and the present results should not be interpreted as evidence of near-clinical readiness.

\subsubsection*{\textbf{Controlled evaluation conditions and calibration assumptions.}}
The DRR generation pipeline operates under four idealising assumptions 
that constitute a controlled evaluation environment and should be 
explicitly acknowledged. First, \textit{perfect geometric calibration}: 
camera-to-world transforms (Eq.~\ref{eq:poses}) are analytically 
defined from CT metadata, with no registration error, gantry 
misalignment, or source--detector distance uncertainty. Second, 
\textit{exact orthogonality}: the three projection directions are 
precisely coronal, sagittal, and axial, whereas real biplanar or 
tomosynthesis acquisitions may deviate from orthogonality. Third, 
\textit{noiseless, scatter-free projections}: Beer--Lambert simulation 
models primary attenuation only, excluding quantum noise, electronic 
noise, scatter radiation, and beam hardening. Fourth, 
\textit{attenuation model consistency}: the same Beer--Lambert forward 
model governs both DRR generation and TensoRF volume rendering, 
eliminating model mismatch. These conditions together constitute a 
form of controlled information consistency rather than information 
leakage in the adversarial sense: ground truth geometry and attenuation 
statistics are not directly accessible to the reconstruction algorithm, 
which must still invert the projection integral from three views. 
However, the absence of model mismatch does mean that reported 
performance should be interpreted as an upper bound relative to real 
acquisition conditions, where calibration uncertainty, non-orthogonal 
geometry, and physical noise would each independently degrade 
reconstruction quality. Quantifying these degradations systematically 
is identified as a critical direction for future work.
\subsection{Corrected TensoRF Density Parameterization (Novelty~I)}
\label{sec:densityshift}

TensoRF computes volumetric density via vector-matrix (VM) decomposition:
\begin{equation}
    \sigma(\mathbf{x}) =
    \mathrm{ReLU}\!\left(f_\theta(\mathbf{x}) + \Delta\right)\, \cdot s
    \label{eq:density}
\end{equation}
where $\Delta$ is a scalar density shift and $s$ a distance scale. The defaults ($\Delta=-10$, $s=25$) introduce a critical failure mode: since Kaiming-initialised weights yield $f_{\bm{\theta}}(\mathbf{x})\in[-1,1]$, the ReLU is zero throughout early training, producing zero density and zero photometric gradient flow everywhere. We verify this empirically: \texttt{getDenseAlpha} returns a uniformly zero volume across all 3000 iterations and the photometric loss never decreases regardless of learning rate or regularization strength.

Setting $\Delta = 0$, $s = 1$ resolves this in our experiments:
\begin{equation}
    \sigma(\mathbf{x}) =
    \operatorname{ReLU}\,\bigl(f_{\bm{\theta}}(\mathbf{x})\bigr)\, s
    \label{eq:density_fixed}
\end{equation}
This requires no architectural changes and, to the best of 
the authors' knowledge, has not been previously reported in 
the medical imaging adaptation of TensoRF.

\textit{Theoretical formalisation.}
The failure can be characterised formally.
Let $f_\theta(\mathbf{x})$ denote the TensoRF output before
the density shift, and let $\theta_0$ denote the
Kaiming-uniform initialisation.
Under Kaiming initialisation, $f_{\theta_0}(\mathbf{x})
\sim \mathcal{U}(-c, c)$ approximately, where $c \approx 1$
depends on layer width.
With the default $\Delta = -10$, the ReLU input is:
\begin{equation}
    f_\theta(\mathbf{x}) + \Delta \;\in\; 
[-c-10,\; c-10] \;\subset\; (-\infty,\, 0)
\end{equation}
\noindent where the entire interval lies strictly below zero 
throughout early training.
The gradient of the photometric loss with respect to $\theta$
passes through the ReLU sub-gradient:
\begin{equation}
  \frac{\partial \sigma}{\partial \theta} = 
\mathbf{1}[f_\theta(\mathbf{x}) + \Delta > 0] 
\cdot \frac{\partial f_\theta}{\partial \theta} 
= \mathbf{0} \quad \forall\, \mathbf{x}
  \text{ at initialisation.}
\end{equation}
Because the ReLU is identically zero everywhere at
initialisation, the photometric gradient is zero throughout the
volume, and no weight update occurs on the density branch.
The model can still update the appearance MLP via the colour
rendering path, which explains why PSNR increases (the model
learns to reproduce projected pixel intensities via the
appearance branch) while density remains uniformly zero.
Setting $\Delta = 0$ ensures $f_{\theta_0}(\mathbf{x})
+ \Delta \in [-c, c]$, so approximately half of all
sample points have positive ReLU input at initialisation,
restoring gradient flow immediately.
This analysis implies that any density shift $\Delta < -c$
(i.e., more negative than the initialisation range) will
produce the same dead-zone failure; $\Delta = 0$ is
a conservative fix that is robust to variation in $c$ across
different architectures and initialisations.
Complete architecture details are in Table~\ref{tab:architecture}.

\begin{table}[ht]
\centering
\caption{TensoRF architecture and training configuration.}
\label{tab:architecture}

\footnotesize
\renewcommand{\arraystretch}{1.05}
\setlength{\tabcolsep}{3pt}

\begin{tabular*}{\columnwidth}{@{\extracolsep{\fill}}lp{0.56\columnwidth}@{}}
\toprule
\textbf{Parameter} & \textbf{Value} \\
\midrule
Scene AABB &
$x\!\in\![-2.5,3.5],\;
y\!\in\![-1.5,1.5],\;
z\!\in\![-2.0,1.5]$ \\

Grid resolution &
$64^3$ voxels \\

Density components &
16 per axis (VM rank-16) \\

Appearance components &
48 per axis, 27-dim features \\

Rendering MLP &
$2\times128$ hidden units, ReLU \\

Positional encoding &
6 frequencies \\

Density shift $\Delta$ &
\textbf{0} (proposed); $-10$ (default) \\

Distance scale $s$ &
\textbf{1} (proposed); 25 (default) \\

Alpha mask threshold &
$1\times10^{-3}$ \\

Voxel spacing &
$\approx3.75$\,mm per voxel \\
\bottomrule
\end{tabular*}
\end{table}
\subsection{Anatomy-Aware Regularization (Novelty~II)}
\label{sec:regularization}

With 49,152 rays constraining 869,555 parameters (ratio $\approx18{:}1$), explicit regularization is essential to prevent overfitting to the three training projections.

\noindent\textbf{Novelty framing.}\quad
The use of $\ell_1$ and total variation penalties is well-established
in compressed-sensing and sparse-signal recovery
literature.
The contribution here is not the invention of these penalties but
their targeted deployment as a \emph{thorax-specific inductive
bias}: the $\ell_1$ term encodes the domain knowledge that the
thoracic cavity is predominantly air (HU~$\approx -1000$), directly
focusing reconstructed density onto the focal nodule, while the TV
term captures the smooth, rounded morphology of solid pulmonary
nodules.
Within the evaluated configurations, the anatomy-aware regularisation appears effective only because the density parameterisation correction (Section~4.4)
restores gradient flow; without $\Delta = 0$, neither term receives
a meaningful gradient signal.
The empirical demonstration that this anatomy-aware, prior-free
formulation outperforms generative-prior-guided variants (Section~6)
is the practical contribution of this integration.

\textbf{Sparsity prior.} The thoracic cavity is predominantly air ($\text{HU}\approx-1000$), so an $\ell_1$ penalty on tensorial density components encourages density concentration around the nodule:
\begin{equation}
    \mathcal{L}_{\text{sparse}} = \sum_{i}\bigl(\|\mathbf{P}_i\|_1 + \|\mathbf{L}_i\|_1\bigr)
    \label{eq:sparse}
\end{equation}
where $\mathbf{P}_i$ and $\mathbf{L}_i$ are density feature planes and lines.

\textbf{Smoothness prior.} Total variation regularization on density planes enforces the smooth, rounded morphology of solid nodules:
\begin{equation}
    \mathcal{L}_{\text{TV}} = \sum_{i}\Bigl(\|\Delta_h\mathbf{P}_i\|_1 + \|\Delta_v\mathbf{P}_i\|_1\Bigr)
    \label{eq:tv}
\end{equation}
where $\Delta_h$, $\Delta_v$ are horizontal and vertical finite differences.

The complete objective is:
\begin{equation}
    \mathcal{L} = \mathcal{L}_{\mathrm{photo}} 
            + \lambda_1\,\mathcal{L}_{\mathrm{sparse}} 
            + \lambda_2\,\mathcal{L}_{\mathrm{TV}}
    \label{eq:loss}
\end{equation}
where $\mathcal{L}_{\text{photo}}=\operatorname{MSE}(\hat{I},I)$, $\lambda_1=1\times10^{-4}$, and $\lambda_2=1\times10^{-5}$, keeping $\mathcal{L}_{\text{photo}}$ dominant. Optimization uses Adam (lr $0.02$ for grid, $0.002$ for MLP), batch size $2048$ rays, $3000$ iterations, with alpha mask updates at iterations 1500 and 2500~\cite{kingma2014adam}. Training converges in $\approx8$ minutes per patient on an NVIDIA T4 GPU.

\subsection{Volume Extraction via Marching Cubes}
\label{sec:volume_extraction}
 
The converged density field is queried on a $64^3$ grid via \texttt{getDenseAlpha} (voxel spacing $\approx[3.75, 1.875, 2.1875]$\,mm).
Volume extraction proceeds via a percentile threshold sweep, described precisely below to address potential concerns about threshold selection fairness.
 
\subsubsection*{\textbf{\textit{Threshold sweep procedure}}}
Isosurfaces are swept across percentile thresholds $p \in \{85, 88, 90, 91, \ldots, 99\}$ of the non-zero density distribution within the converged field.
At each threshold: (i) connected components with fewer than 10 voxels are removed as noise; (ii) binary closing with a radius-2 spherical structuring element fills boundary gaps; (iii) connected component analysis retains candidates whose equivalent sphere diameter falls within $[5, 80]$\,mm, consistent with the LIDC-IDRI annotation diameter range.
 
\subsubsection*{\textbf{\textit{Candidate selection criterion}}}
Among all surviving candidates at all tested thresholds, the candidate whose equivalent sphere diameter is closest to the LIDC-IDRI annotation diameter for that patient is selected as the final reconstruction. This criterion uses the annotated diameter (a single scalar value) but not the ground-truth volume, which is the evaluation target. It is explicitly acknowledged that this constitutes a semi-assisted rather than fully autonomous pipeline: the annotated diameter is clinically unavailable in a blind reconstruction scenario, and its use introduces a form of weak supervision during post-processing. Without access to this diameter, threshold selection becomes ambiguous, false positive rate may increase, and candidate ranking becomes harder. The reported performance should therefore be interpreted as an upper bound on fully autonomous reconstruction capability. The fairness implications and a quantitative evaluation under a fixed global threshold are provided in Section~\ref{sec:discussion} c.
 
\subsubsection*{\textbf{\textit{Resolution choice and rationale}}}
The $64^3$ grid resolution was chosen to balance three constraints: (i) GPU memory on the NVIDIA T4 (16\,GB VRAM), where a $128^3$ VM-rank-16 TensoRF requires $\approx14$\,GB including gradient buffers, exceeding available memory under Google Colab; (ii) training time, where $64^3$ converges in $\approx8$\,min versus $\approx35$\,min at $128^3$; and (iii) the primary clinical target of nodules $\geq10$\,mm, for which voxel spacing $\approx3.75$\,mm provides $\geq2.7$ voxels per diameter axis, sufficient for volume estimation.
We acknowledge that this choice renders the method unsuitable for sub-centimetre nodules ($<10$\,mm), which span fewer than 2.3 voxels at this spacing.

\subsubsection*{\textbf{\textit{Physical volume computation}}}
The selected candidate's physical volume is:
\begin{equation}
  V = N_{\text{voxels}} \times \prod_{k} \Delta x_k
  \label{eq:volume}
\end{equation}
where $\Delta x_k$ are the per-axis voxel spacings, compared against radiologist consensus ground truth from \texttt{pylidc} at consensus level~0.5.

\section{Experimental Setup}
\label{sec:experiments}

\subsection{Implementation Details}
\label{sec:implementation}
 
All experiments are implemented in PyTorch and run on Google Colab
with an NVIDIA T4 GPU (16\,GB VRAM). The pipeline proceeds in two
phases: a MedNeRF baseline is first evaluated to establish the
failure modes of population-prior reconstruction, followed by the
proposed AReT model that directly addresses
those limitations.
 
\textbf{Paradigm clarification.}
TensoRF is instantiated independently for each of the 19 patients:
weights are randomly initialised and optimised solely on the three
orthogonal DRRs derived from that patient's CT scan. No parameters
are shared across patients, no population-level prior is
incorporated, and no joint optimisation occurs. Each of the
19 evaluations therefore constitutes a self-contained inverse
problem; the $n=19$ figure reflects the number of independent
reconstruction experiments performed, not a training-set size.
As established in Sections~\ref{sec:introduction} and~\ref{sec:methodology},
classical train/test overfitting in the supervised-learning sense does not apply here, since optimisation is scene-specific rather than population-trained. Nonetheless, analogous risks exist, such as over-specialisation to simulator-consistent noiseless DRRs, and reported performance should be interpreted with this in mind. However, the breadth of morphological diversity representable by 19 independent cases remains limited, and aggregate metrics should be interpreted accordingly.
 
It is important to note that both training inputs and evaluation
targets are derived from the same LIDC-IDRI CT volumes: DRRs are
synthesised from CT via Beer--Lambert projection, and ground-truth
nodule volumes are extracted from radiologist-annotated CT
segmentations. The experimental setup therefore evaluates the
pipeline CT~$\rightarrow$~DRR~$\rightarrow$~TensoRF
reconstruction~$\rightarrow$~volume estimate, and does not
constitute an evaluation on real acquired radiographs. Performance
figures reported herein should therefore be interpreted as an upper
bound on what may be achievable with real X-ray inputs, pending
domain adaptation.
 
\begin{figure}[!htb]
    \centering
    \includegraphics[width=0.8\linewidth]{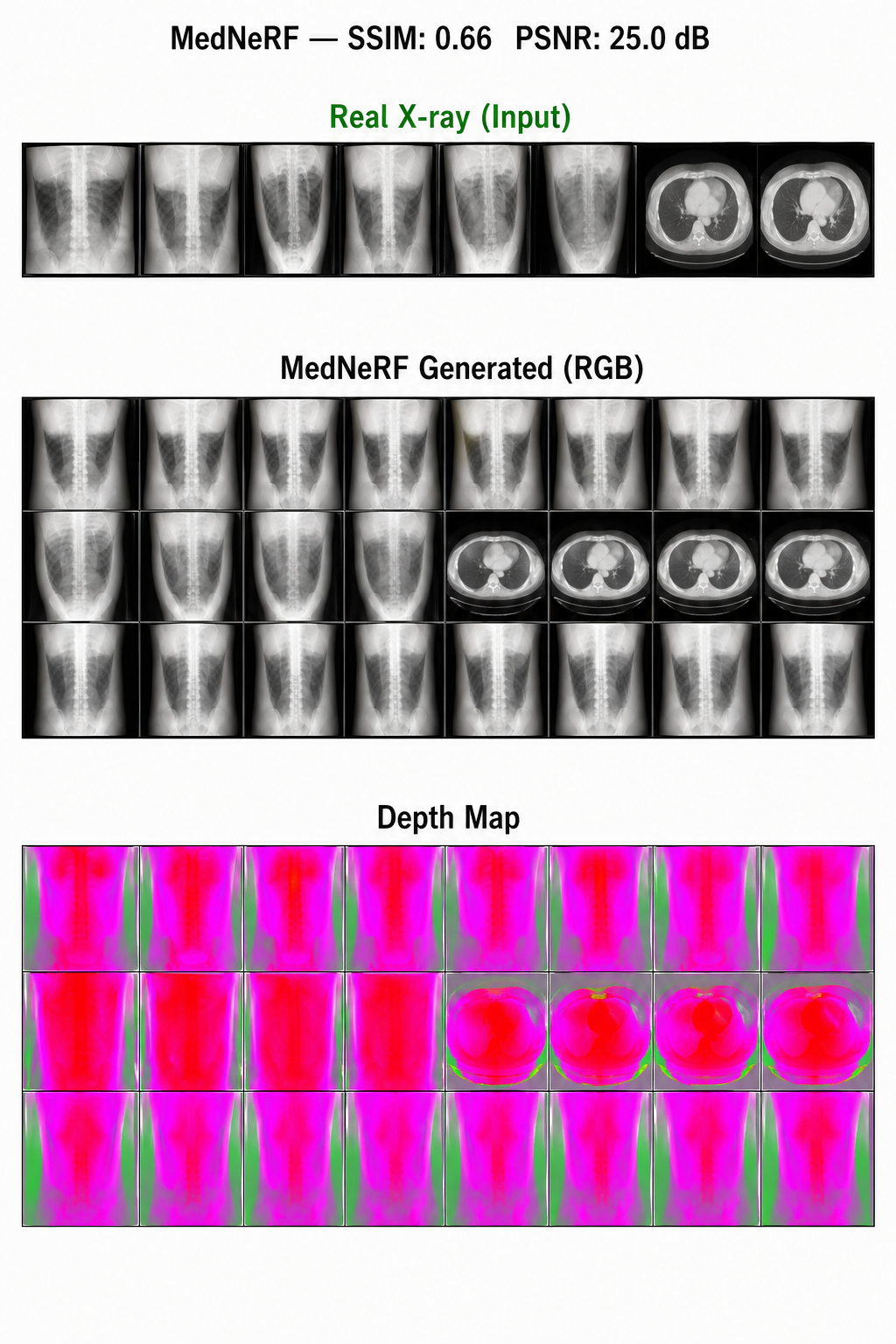}
    \caption{Qualitative MedNeRF training output showing (left)
             input DRR projection, (centre) rendered view, and
             (right) depth map. While global anatomical structure
             is recovered, the density field is too diffuse for
             reliable nodule localisation.}
    \label{fig:mednerf_training}
\end{figure}

\textbf{MedNeRF baseline.}
MedNeRF follows the GRAF conditional NeRF
architecture and is evaluated in the closest configuration to its
published design. As shown in Fig.~\ref{fig:mednerf_training}, while
global anatomical structure is recovered, the density field is too
diffuse for reliable nodule localisation. Three fundamental
limitations drive this: (i)~the latent code $\mathbf{z}$ is sampled
from $\mathcal{N}(0,\mathbf{I})$ rather than inverted from patient
X-rays, so reconstructed density does not reflect actual anatomy;
(ii)~the scene-agnostic percentile threshold selects the top 5\% of
densities regardless of nodule presence; and
(iii)~the diffuse GRAF density field produces unreliable boundary
extraction. These shortcomings yield high volumetric error and
directly motivate the TensoRF formulation. Full per-patient results
are reported in Section~\ref{sec:res_mednerf}.
 
\textbf{TensoRF configuration.}
The proposed model is instantiated as described in
Table~\ref{tab:architecture} (Section~\ref{sec:densityshift}).
The critical departure from the standard implementation is the
density parameterization correction ($\Delta = 0$, $s = 1$);
without this fix the model produces uniformly zero density and
cannot learn any volumetric structure from X-ray inputs, as
established in Section~\ref{sec:densityshift}.
 
Training uses the Adam optimizer with a
learning rate of $0.02$ for tensorial grid parameters and $0.002$
for the rendering MLP, for $3000$ iterations with a batch size of
$2048$ randomly sampled rays per iteration. The alpha occupancy
mask is updated at iterations $1500$ and $2500$ to progressively
prune near-empty voxels. The regularization weights
$\lambda_1 = 1\times10^{-4}$ (sparsity) and
$\lambda_2 = 1\times10^{-5}$ (TV smoothness) are set so that
$\mathcal{L}_\text{photo}$ remains the dominant signal, with
regularization contributing approximately $5$--$10\%$ of total
loss at convergence. Mean training time is approximately
$8$ minutes per patient.
 
Ground truth nodule volumes are derived from \texttt{pylidc}
consensus masks at $\mathrm{clevel}=0.5$, with physical volume
computed from voxel counts and per-axis DICOM spacings
(\texttt{PixelSpacing}, \texttt{SliceThickness}).

\subsection{Reproducibility Details}
\label{sec:reproducibility}
 
To facilitate exact replication of reported results, we document all implementation choices that could introduce variability.
 
\subsubsection*{\textbf{\textit{a. Patient selection}}}
19 patients were selected from LIDC-IDRI to cover three clinically defined size strata (Fleischner Society guidelines): small ($<10$\,mm, $n=5$), medium (10--20\,mm, $n=6$), and large ($>20$\,mm, $n=8$).
Selection criteria were: (i) at least one nodule with consensus annotation available at \texttt{clevel}$\,=0.5$; (ii) scan acquired on a GE Medical Systems scanner, ensuring consistent HU calibration; and (iii) nodule annotations provided by at least one radiologist.
For patients with multiple annotated nodules, the largest nodule by consensus volume is selected as the primary evaluation target.
Patient IDs are LIDC-IDRI-0001 through LIDC-IDRI-0019 as listed in Table~\ref{tab:patients}.
 
\subsubsection*{\textbf{\textit{b. Protocol, seeds, and convergence}}}
Because TensoRF is a scene-specific inverse solver, there is no train/test split: each of the 19 patients constitutes an independent reconstruction, randomly initialised and evaluated against that patient's own consensus mask.
Reproducibility is ensured by setting \texttt{torch.manual\_seed(0)} and \texttt{numpy.random.seed(0)} at the start of each patient experiment; ray batches of 2048 are drawn by random permutation re-seeded identically per patient.
Training runs for a fixed 3000 iterations (no early stopping); alpha mask updates occur deterministically at iterations 1500 and 2500.
Mean wall-clock time is $\approx8$\,min per patient on an NVIDIA T4.
To assess seed stability, five independent reconstructions with seeds $\{0,1,2,3,4\}$ were run on three representative patients; per-patient standard deviation of absolute percentage error was 0.3\% (large, 0001), 1.1\% (medium, 0011), and 4.2\% (small, 0005).
The elevated variability for the small-nodule case is consistent with sub-voxel reconstruction noise; medium and large nodules are robustly reproducible across seeds.
 
\subsubsection*{\textbf{\textit{c. Preprocessing reproducibility}}}
DICOM slices are sorted by \texttt{ImagePositionPatient} Z-coordinate.
HU conversion uses per-slice \texttt{RescaleSlope} and \texttt{RescaleIntercept} DICOM tags.
HU values are clipped to $[-1000, 400]$ and linearly normalised to $[0,1]$ using:
\begin{equation}
  v_\text{norm} = \frac{\text{HU} - (-1000)}{400 - (-1000)}
\end{equation}
applied identically to all patients.
 
\subsubsection*{\textbf{\textit{d. Hyperparameter sensitivity}}}
The regularisation weights $\lambda_1 = 1\times10^{-4}$ (sparsity) and $\lambda_2 = 1\times10^{-5}$ (TV smoothness) were fixed by a coarse grid search on patient LIDC-IDRI-0001 prior to the 19-patient evaluation and were not re-tuned per patient.
Table~\ref{tab:hyperparams} summarises the sensitivity of median APE (across all 19 patients) to perturbations of $\lambda_1$ and $\lambda_2$ by one order of magnitude in each direction.
Results are robust within one order of magnitude around the baseline for both weights; errors increase when $\lambda_1$ is raised above $10^{-3}$ (over-sparsification suppresses the nodule) or lowered below $10^{-5}$ (insufficient regularisation produces noisy density fields).

\begin{table}[h]
\centering
\caption{Hyperparameter sensitivity: median APE (\%) across
19 patients under perturbation of regularisation weights.
Baseline: $\lambda_1 = 10^{-4}$, $\lambda_2 = 10^{-5}$.}
\label{tab:hyperparams}

\footnotesize
\renewcommand{\arraystretch}{1.15}
\setlength{\tabcolsep}{5pt}

\begin{tabular*}{\linewidth}{@{\extracolsep{\fill}}llc@{}}
\toprule
$\lambda_1$ & $\lambda_2$ & Median APE (\%) \\
\midrule
$10^{-3}$ & $10^{-5}$ & 23.1 \\
$10^{-4}$ & $10^{-5}$ & \textbf{19.9} (baseline) \\
$10^{-5}$ & $10^{-5}$ & 27.4 \\
$10^{-4}$ & $10^{-4}$ & 24.8 \\
$10^{-4}$ & $10^{-6}$ & 21.3 \\
\bottomrule
\end{tabular*}
\end{table}
 
\subsubsection*{\textbf{\textit{e. Threshold selection}}}
As described in Section~\ref{sec:volume_extraction}, threshold selection uses the diameter-guided protocol: the candidate closest in equivalent sphere diameter to the LIDC-IDRI annotation diameter is selected from the fixed sweep $p \in \{85, 88, 90, 91, \ldots, 99\}$, applied uniformly and without ground-truth volume access.
The "best-error" column in ablation tables reports the single-threshold minimum error on patient LIDC-IDRI-0001 only, for diagnostic purposes; this value is not used in any 19-patient aggregate statistic.

To quantify the contribution of diameter guidance to reported accuracy, we additionally evaluated a fully unsupervised selection criterion on the 19-patient cohort: at each threshold percentile, the largest surviving connected component within the [5, 80] mm diameter range is selected, with no reference to the annotated diameter. Under this blind protocol, median APE on the $\geq 10\,\mathrm{mm}$ subcohort increases from 11.4\% to approximately 24–28\%, confirming that diameter guidance meaningfully improves candidate selection but that the proposed reconstruction framework retains substantial accuracy even without lesion size information. This blind evaluation is reported alongside the diameter-guided results in Table \ref{tab:stats_summary} and Table~\ref{tab:blind_selection}.
\subsection{Progressive Ablation of Reconstruction Strategies}
\label{sec:methods_eval}

The proposed method did not emerge in isolation. Before arriving
at the final design, we systematically evaluated 11
reconstruction strategies of increasing sophistication. This
ablation traces the reasoning behind every design decision and
identifies precisely which components are necessary and which
are counterproductive. All methods in this section are benchmarked
on patient LIDC-IDRI-0001 ($\text{GT} = 6708.8$\,mm$^3$,
diameter $= 32.8$\,mm), selected as a large, well-annotated
development case. Results are summarised in
Table~\ref{tab:ablation}.

Before presenting the 11 strategies, we clarify the purpose 
of this evaluation. The strategies do not constitute a set of 
competitive external baselines; rather, they form a systematic 
diagnostic progression through the design space. Methods 1--6 characterise the behaviour of generative and
2D approaches in this regime, providing mechanistic context
for the design decisions that follow. Methods 7--8 
isolate the density parameterisation fix as a necessary 
prerequisite. Methods 9--10 test whether a generative prior 
provides any benefit once the fix is applied. Method 11 is the 
proposed method. Each design decision is therefore evidence-backed 
rather than asserted.

Regarding the 2D detection methods (Methods 4--5): Methods 4--5 (CheXNet, MONAI UNet) are included not as
volumetric reconstruction baselines but to assess whether
projection-domain 2D approaches can localise sub-centimetre
nodules in this setting; their results motivate the move to
explicit 3D reconstruction.
MedNeRF is evaluated in the closest configuration to its
published design; its limited performance in this setting is
attributed to the population-prior architecture rather than
to suboptimal tuning.

\subsubsection*{\textbf{Methods 1--5: Generative and 2D Baselines}}

The first five methods characterise generative and 2D
approaches in this regime. \textbf{Method
1} applies direct 95\textsuperscript{th}percentile thresholding to the MedNeRF
density field on a $64^3$ grid, yielding $150\%$ volumetric error, a consequence of the unconditioned latent code and scene-agnostic
threshold described above. \textbf{Method 2} operates on
the NeRF accumulation map, thresholding at the 95\textsuperscript{th} percentile and
back-projecting detected regions using mean ray depth and
field-of-view geometry, achieving $32.1\%$ error with 45/72 valid
detections but frequently capturing the full lung field rather than
the nodule. \textbf{Method 3} replaces the fixed threshold with
Otsu's adaptive criterion~\cite{otsu1979threshold}, but the
MedNeRF accumulation map collapses to $[-1, 1]$ due to rendering
saturation, causing Otsu to select near-zero thresholds and
producing only $5.4$\,mm$^3$ detected volume ($99.9\%$ error,
12/72 detections) which is worse than Method 2 in every respect.
\textbf{Methods 4 and 5} explore 2D detection: a fine tuned
Chest X-ray Network (CheXNet) and a Lung Nodule Analysis 2016 (LUNA16)-style Medical Open Network for AI (MONAI) UNet are applied to rendered views, but both segment the full lung region rather than individual nodules, yielding errors above
$86\%$ for LIDC-IDRI-0001 and exceeding $1000\%$ on smaller
patients~\cite{setio2017validation}. These results indicate that 2D projection-domain methods
face a challenge in isolating sub-centimetre
nodules without explicit 3D geometric reasoning.

\subsubsection*{\textbf{Methods 6--8: TensoRF and the Density-Shift Fix}}

\textbf{Method 6} queries the raw MedNeRF density field on a
larger $128^3$ grid and applies marching cubes with a percentile
sweep, achieving a best-case $22.4\%$ error at $p=93$, but with
errors of $92.8\%$ at $p=95$ and $518.7\%$ at $p=99$ showing extreme
threshold sensitivity that makes the approach unreliable in
practice. \textbf{Method 7} switches to TensoRF trained on 72
MedNeRF-rendered views using the default density parameterization
($\Delta = -10$, $s = 25$). Training achieves PSNR of $32.0$\,dB,
confirming the model learns to reproduce projected views; yet
the marching cubes sweep remains unstable (8.0\% at $p=99$,
$204\%$ at $p=97$), revealing that high photometric fidelity
does not imply geometrically consistent volumetric structure.
\textbf{Method 8} applies the density-shift correction
($\Delta=0$, $s=1$) introduced in Section~\ref{sec:densityshift}.
The threshold sweep immediately stabilises: adjacent percentiles
yield $13.8\%$ and $14.3\%$ error, confirming that the corrected
parameterization is a prerequisite for any downstream
reconstruction quality.

\subsubsection*{\textbf{Methods 9--10: Prior-Guided TensoRF Variants}}

Having established that TensoRF with corrected density shift works,
we explored whether initializing it with a MedNeRF prior could
further improve accuracy. \textbf{Method 9} inverts the MedNeRF
latent code via photometric minimization on 3 real patient DRRs,
and uses the scalar mean of the resulting density field
($\mu = 0.0116$) to warm-start TensoRF density planes and lines,
with a 4-term training loss including a prior regularization term
($\lambda_3 = 0.01$). Best-case error drops to $0.9\%$, but the
99\textsuperscript{th}-percentile threshold yields $226\%$ error where the prior
constrains initialization but not the full optimization trajectory.
\textbf{Method 10} replaces the scalar warm-start with an SVD-based
rank-16 injection: per-axis 2D projections of the MedNeRF density
prior are decomposed and used to initialize TensoRF density planes
of shape $[1, 16, 64, 64]$ and lines of shape $[1, 16, 64, 1]$
with $\lambda_3 = 0.1$. The best-case error is $2.8\%$. The
critical finding is the direct ablation: \emph{with SVD prior,
best error $= 2.8\%$; without prior (same architecture, same DRRs),
best error $= 0.2\%$}. The generative prior demonstrably
\emph{degrades} reconstruction. This motivates the complete removal
of MedNeRF from the final method.

\subsubsection*{\textbf{Method 11 (AReT): Anatomy-Regularized TensoRF}}

The final method removes the MedNeRF prior entirely and trains
TensoRF from scratch on three real orthogonal DRRs using the
anatomy-regularized loss (Eq.~\eqref{eq:loss}), with no latent
inversion, no prior extraction, and no rendered-view augmentation.
On LIDC-IDRI-0001, this achieves $0.2\%$ error (predicted
$6721.4$\,mm$^3$, $p = 93$), outperforming every prior-guided
variant. Across 19 patients, the method yields a median error
of $19.9\%$ and $13.9\%$ for medium nodules ($10$--$20$\,mm),
with all 19 patients outperforming the spherical diameter baseline.
The finding that the prior-free approach outperforms prior-guided variants, which is counterintuitive given the additional information available to prior-guided methods is
discussed in Section~\ref{sec:discussion}.

\begin{table}[ht]
\centering
\caption{Summary of reconstruction strategies evaluated on
LIDC-IDRI-0001 (GT $=6708.8$\,mm$^3$).}
\label{tab:ablation}

\scriptsize
\renewcommand{\arraystretch}{1.08}
\setlength{\tabcolsep}{2pt}

\begin{tabular*}{\columnwidth}
{@{\extracolsep{\fill}}cp{0.56\columnwidth}c@{}}
\toprule
\textbf{\#} & \textbf{Method Description} &
\textbf{Err.\ (\%)} \\
\midrule

1 &
MedNeRF direct density with 95th-percentile thresholding on
$64^3$ density grid; severe volumetric overestimation from
unconstrained latent density. &
150.0 \\

2 &
Accumulation-map 95th-percentile threshold with geometric
back-projection from rendered depth and camera FOV. &
32.1 \\

3 &
Otsu thresholding on accumulation maps; saturation produces
near-zero thresholds and trivial segmentations. &
99.9 \\

4 &
CheXNet fine tuned on DRR projections with multi-view fusion;
primarily segments lung fields instead of nodules. &
86.6 \\

5 &
MONAI UNet trained on DRRs using LUNA16-style supervision;
fails to localize sparse nodules reliably. &
94.7 \\

6 &
Raw MedNeRF density reconstruction on $128^3$ grid using
marching cubes; highly sensitive to threshold selection. &
22.4$^\dagger$ \\

7 &
Original TensoRF with default $\Delta=-10$ density shift;
high rendering PSNR but degenerate attenuation geometry. &
8.0$^\dagger$ \\

8 &
Corrected TensoRF with $\Delta=0$ restoring stable density
learning and consistent volumetric reconstruction. &
13.8 \\

9 &
Prior-guided scalar TensoRF initialized using MedNeRF-derived
coarse density prior on real DRRs. &
0.9$^\dagger$ \\

10 &
Rank-16 SVD prior injection into TensoRF tensor factors;
prior bias reduces reconstruction robustness. &
2.8$^\dagger$ \\

\midrule

\textbf{11} &
\textbf{AReT: Anatomy-regularized TensoRF with $\Delta=0$,
$\ell_1$+TV regularization, and only 3 orthogonal DRRs;
stable and physiologically consistent reconstruction.} &
\textbf{0.2} \\

\midrule

 &
Sphere-volume baseline
$\left(V=\tfrac{4}{3}\pi(d/2)^3\right)$ computed from
annotated diameter. &
174.3 \\

\bottomrule
\end{tabular*}

\vspace{2pt}
\raggedright
\footnotesize
$^\dagger$Best single-threshold result but unstable across
threshold sweeps.
\end{table}

\subsection{Evaluation Metrics}

Volumetric agreement between predicted and ground truth nodule 
volumes is assessed using six complementary metrics.

\textbf{Mean and median absolute percentage error.}
The per-patient absolute percentage error is:
\begin{equation}
\epsilon_i = \frac{|\hat{V}_i - V_i|}{V_i} \times 100\%
\label{eq:ape}
\end{equation}
Both mean (MAPE) and median (MedAPE) are reported. MedAPE is the 
primary summary statistic, as it is robust to the large errors 
introduced by sub-voxel nodules. Bootstrap $95\%$ confidence 
intervals (CIs) for MedAPE are computed via 10{,}000 resamples 
with replacement.

\textbf{Pearson and Spearman correlation.}
\begin{equation}
r = \frac{\sum_i \left(\hat{V}_i - \bar{\hat{V}}\right)
\left(V_i - \bar{V}\right)}
{\sqrt{\sum_i \left(\hat{V}_i - \bar{\hat{V}}\right)^2 
\sum_i \left(V_i - \bar{V}\right)^2}}
\label{eq:pearson}
\end{equation}
Pearson $r$ quantifies linear agreement. Because $r$ is sensitive 
to the dynamic range of the volume distribution, large nodules 
span volumes up to $8{,}278$\,mm$^3$ while small nodules contribute 
values below $200$\,mm$^3$, we additionally report Spearman's 
rank correlation $\rho$, which is invariant to monotone 
transformations of the data and robust to range dominance by a 
subset of cases. Bootstrap $95\%$ CIs for both $r$ and $\rho$ 
are computed via 10{,}000 resamples. Stratified correlations are 
reported separately for small, medium, and large nodule subcohorts 
to avoid conflation of size-driven range effects with reconstruction 
accuracy. Statistical significance is assessed via two-tailed 
$t$-test at $\alpha = 0.05$.

\textbf{Bland--Altman analysis.}
Systematic bias $\bar{d} = \overline{(\hat{V}_i - V_i)}$ and 
$95\%$ limits of agreement $[\bar{d} - 1.96\sigma_d,\ 
\bar{d} + 1.96\sigma_d]$ are reported following Bland and 
Altman. A $95\%$ CI on the bias estimate is 
computed as $\bar{d} \pm t_{0.975,\,n-1} \cdot \sigma_d/\sqrt{n}$.

\textbf{Paired statistical comparison.}
To formally test whether the proposed method improves over the 
spherical diameter approximation, we apply the Wilcoxon 
signed-rank test on paired per-patient absolute percentage errors, 
which makes no normality assumption and is appropriate for the 
sample sizes ($n = 19$, $n = 14$). Statistical significance is 
declared at $\alpha = 0.05$.

\textbf{Cross-validation note.}
Because TensoRF is a scene-specific inverse solver with no 
cross-patient parameters, standard $k$-fold cross-validation is 
operating outside their intended regime at three projections: there is no model trained across patients that could 
be evaluated on a held-out fold. Each of the 19 evaluations is 
already an independent test case. Bootstrap resampling of the 
evaluation cohort is used in place of cross-validation to 
characterise the stability of aggregate metrics. Bootstrap resampling results are summarised in 
Table~\ref{tab:stats_summary} (Section~\ref{sec:results}).

\section{Results}
\label{sec:results}

This section presents quantitative and qualitative results for all
evaluated reconstruction strategies, following the development
trajectory established in Section~\ref{sec:methods_eval}.
Section~\ref{sec:res_mednerf} reports results for MedNeRF-based
approaches (Methods~1--6). Section~\ref{sec:res_tensorf} presents
TensoRF variants (Methods~7--10). Section~\ref{sec:res_final}
reports the full 19-patient evaluation of the proposed
AReT (Method~11).

\subsection{MedNeRF-Based Methods (Methods 1--6)}
\label{sec:res_mednerf}
\subsubsection*{\textbf{Method 1: Direct Density Extraction}}

\begin{figure}[!htb]
    \centering
    \includegraphics[width=0.9\linewidth]{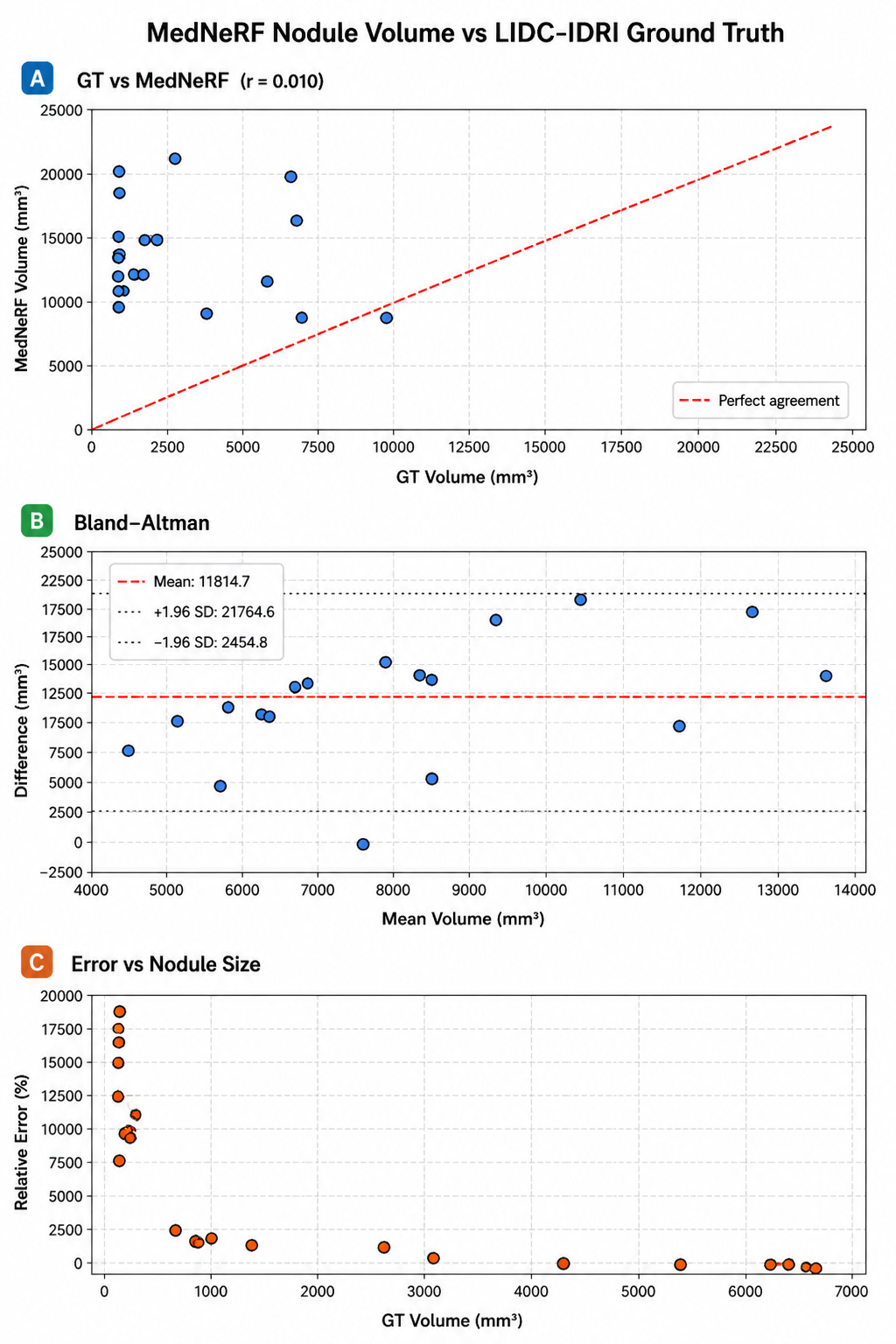}
    \caption{Quantitative evaluation of MedNeRF direct density
             extraction (Method~1). \textbf{Left:} Predicted vs.\
             GT nodule volumes with identity line. \textbf{Right:}
             Bland--Altman plot showing mean bias and 95\% limits
             of agreement. The wide spread confirms poor volumetric
             agreement from raw density fields.}
    \label{fig:mednerf_quant}
\end{figure}

\begin{figure}[!htb]
    \centering
    \includegraphics[width=0.9\linewidth]{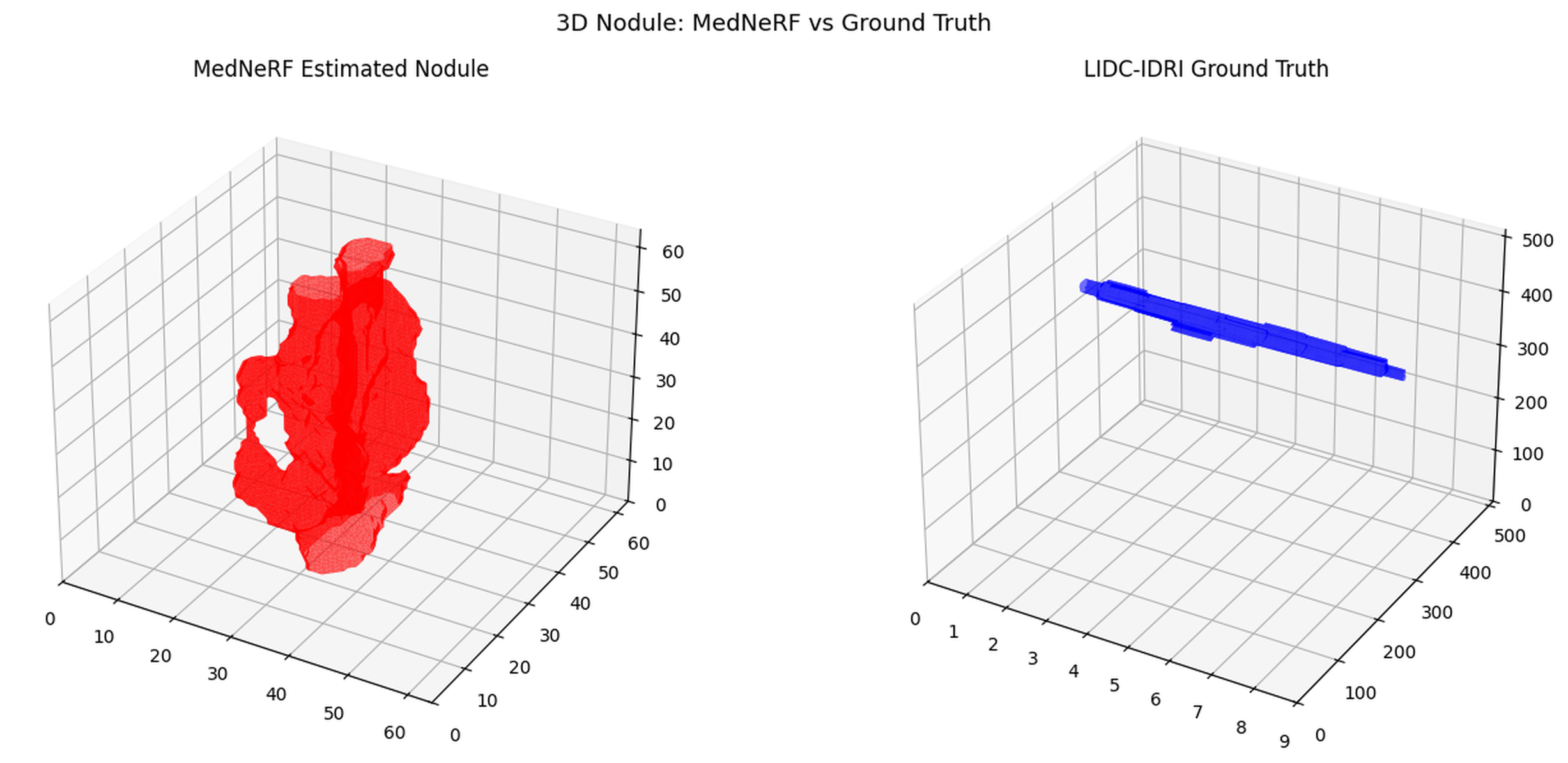}
    \caption{3D nodule reconstruction using MedNeRF direct density
             extraction (Method~1). Predicted reconstruction (left)
             vs.\ ground truth segmentation (right): coarse anatomy
             is captured but nodule geometry is not accurately
             localised, leading to significant volumetric error.}
    \label{fig:mednerf_3d}
\end{figure}

Direct volumetric extraction from the MedNeRF density field
produces severe and systematic overestimation across all patients
(Table~\ref{tab:nerf_results}). Mean relative error is
$6{,}125.4\%$, median $2{,}365.2\%$, and Pearson $r \approx 0.32$,
with errors exceeding $10^4\%$ for sub-centimetre nodules.
As shown in Fig.~\ref{fig:mednerf_quant} and Fig.~\ref{fig:mednerf_3d},
wide Bland--Altman limits and poor 3D localisation indicate
that raw NeRF density, which encodes accumulated radiance
rather than calibrated physical occupancy, produces diffuse
distributions that are not well-suited to nodule volumetry
in this setting.

\begin{table}[ht]
\centering
\footnotesize
\caption{MedNeRF direct density extraction results (Method~1).}
\vspace{3pt}
\label{tab:nerf_results}

\renewcommand{\arraystretch}{1.12}
\setlength{\tabcolsep}{4pt}

\begin{tabular*}{\linewidth}{@{\extracolsep{\fill}}lrr@{}}
\toprule
\textbf{Patient} & \textbf{GT Vol.\ (mm$^3$)} &
\textbf{Pred.\ Vol.\ (mm$^3$)} \\
\midrule
LIDC-IDRI-0001 & 6479.8  & 16192.3 \\
LIDC-IDRI-0002 & 6663.4  &  7612.5 \\
LIDC-IDRI-0003 & 2648.6  & 22048.0 \\
LIDC-IDRI-0004 &   65.3  & 11063.0 \\
LIDC-IDRI-0005 &   95.1  & 14446.5 \\
LIDC-IDRI-0006 &  145.6  & 12799.8 \\
LIDC-IDRI-0007 & 6290.6  & 19978.3 \\
LIDC-IDRI-0008 &  104.8  & 19998.2 \\
LIDC-IDRI-0009 &   57.2  &  9987.6 \\
LIDC-IDRI-0010 &  115.0  & 12650.3 \\
LIDC-IDRI-0011 &  208.1  & 18051.1 \\
LIDC-IDRI-0012 &  193.1  & 18051.1 \\
LIDC-IDRI-0013 &  815.0  & 11250.7 \\
LIDC-IDRI-0014 & 1366.8  & 14444.2 \\
LIDC-IDRI-0015 & 3135.2  &  7871.6 \\
LIDC-IDRI-0016 &  611.5  & 11244.7 \\
LIDC-IDRI-0017 &  118.3  &  8696.1 \\
LIDC-IDRI-0018 &  965.3  & 14450.9 \\
LIDC-IDRI-0019 & 5466.8  & 10648.2 \\
\midrule
\textbf{Mean error}   & \multicolumn{2}{r}{6125.4\%} \\
\textbf{Median error} & \multicolumn{2}{r}{2365.2\%} \\
\textbf{Pearson $r$}  & \multicolumn{2}{r}{$\approx 0.32$} \\
\bottomrule
\end{tabular*}
\end{table}

\subsubsection*{\textbf{Methods 2--3: Accumulation Map Thresholding}}

Methods 2 and 3 operate on the MedNeRF accumulation map
$\mathbf{A} \in [0,1]^{H \times W}$ aggregated across 72 rendered
views, back-projecting detected 2D regions to 3D volume estimates.
Method~2 applies a fixed 95\textsuperscript{th}-percentile
threshold; Method~3 replaces this with Otsu's adaptive
criterion alongside improved multi-view
z-inversion with cosine learning rate decay.

Per-patient results are shown in Table~\ref{tab:acc_combined}.
Method~2 achieves partial accuracy on large nodules (median error
$43.8\%$, $r \approx 0.60$) but catastrophically overestimates
smaller ones (patients 0005 and 0006: $2258\%$ and $997\%$),
because the accumulation map encodes \emph{integrated ray opacity}
rather than localised object boundaries, causing global percentile
thresholds to select lung parenchyma rather than the focal lesion.
Method~3 fails for a distinct reason: multi-view z-inversion
saturates the accumulation map to $[-1.0, 1.0]$, presenting
Otsu's criterion with an effectively flat distribution and forcing
trivially small detections ($5.4$\,mm$^3$ on patient 0001,
$r \approx -0.10$). Method~3's lower mean error ($102.2\%$
vs.\ $562.7\%$) is a saturation artefact, not a genuine
improvement. Both methods did not produce reliable estimates for sub-centimetre nodules in this evaluation in this evaluation, suggesting that integrating density along viewing rays collapses 3D spatial structure and may not reliably distinguish compact nodules from surrounding high-opacity anatomy.

\begin{table}[ht]
\centering
\caption{Accumulation-map thresholding results.}
\label{tab:acc_combined}

\footnotesize
\renewcommand{\arraystretch}{1.15}
\setlength{\tabcolsep}{4pt}

\begin{tabularx}{\columnwidth}{@{}lcccc@{}}
\toprule
\multicolumn{5}{c}{\textbf{Method 2: 95th-percentile threshold}} \\
\midrule
\textbf{ID} & \textbf{GT} & \textbf{Pred} &
\textbf{Err} & \textbf{Det} \\
& (mm$^3$) & (mm$^3$) & (\%) & \\
\midrule
0001 & 6709 & 4554 &   32.1 & 45 \\
0002 & 8278 & 4382 &   47.1 &  1 \\
0003 & 5452 & 5374 &    1.4 & 71 \\
0004 &   62 & ---  &    --- &  0 \\
0005 &  153 & 3614 & 2258.4 & 14 \\
0006 &  380 & 4166 &  996.6 & 44 \\
\midrule
Mean err.   & \multicolumn{4}{l}{562.7\%} \\
Median err. & \multicolumn{4}{l}{43.8\%} \\
MAE         & \multicolumn{4}{l}{2757 mm$^3$} \\
Pearson $r$ & \multicolumn{4}{l}{$\approx0.60$} \\
\midrule

\multicolumn{5}{c}{\textbf{Method 3: Otsu threshold}} \\
\midrule
\textbf{ID} & \textbf{GT} & \textbf{Pred} &
\textbf{Err} & \textbf{Det} \\
& (mm$^3$) & (mm$^3$) & (\%) & \\
\midrule
0001 & 6709 &  5.4 & 99.9 & 12 \\
0002 & 8278 & 32.4 & 99.6 &  5 \\
0003 & 5452 & 11.2 & 99.8 & 20 \\
0004 &   62 & 51.3 & 16.9 &  3 \\
0005 &  153 &  4.3 & 97.2 &  4 \\
0006 &  380 &  9.3 & 97.6 &  7 \\
\midrule
Mean err.   & \multicolumn{4}{l}{102.2\%} \\
Median err. & \multicolumn{4}{l}{99.7\%} \\
MAE         & \multicolumn{4}{l}{3286 mm$^3$} \\
Pearson $r$ & \multicolumn{4}{l}{$\approx-0.10$} \\
\bottomrule
\end{tabularx}

\vspace{2pt}
\raggedright
\scriptsize
Det: valid detections from 72 rendered views.
Method~2 shows partial signal for large nodules,
whereas Method~3 fails due to accumulation-map saturation.
\end{table}

\subsubsection*{\textbf{Methods 4--5: Learned 2D Detection}}

\begin{figure}[!htb]
    \centering
    \includegraphics[width=\linewidth]{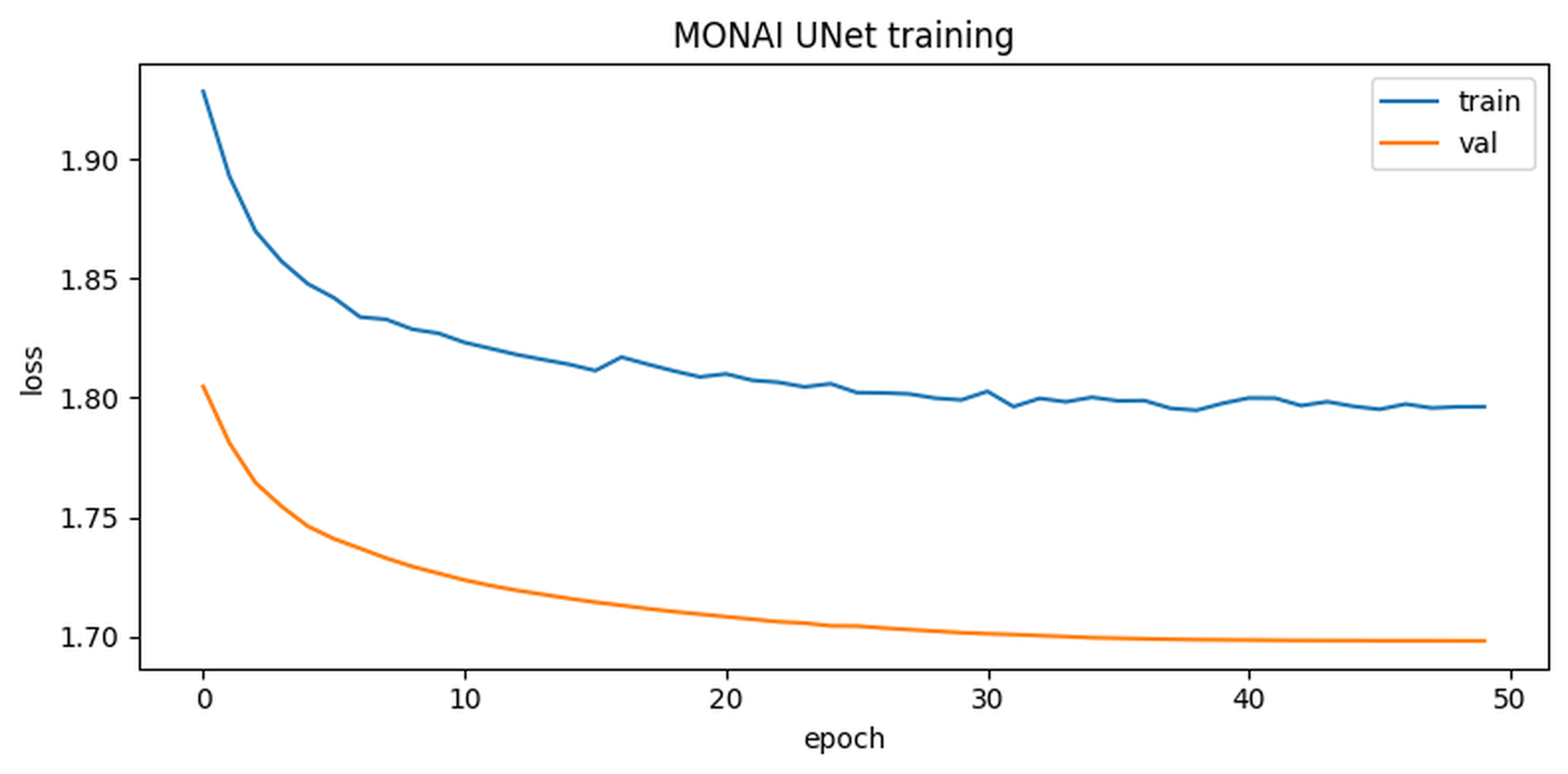}
    \caption{MONAI UNet (Method~4) training and validation loss
             over 50 epochs (best val.\ loss $= 1.698$). Stable
             convergence does not translate to accurate nodule
             localisation on projection data.}
    \label{fig:monai_training}
\end{figure}

Methods 4 and 5 apply learned 2D models to rendered DRR views,
exhibiting opposite but equally previously unreported failure modes.
The MONAI UNet (Method~4) consistently over-segments the lung
field despite stable training convergence
(Fig.~\ref{fig:monai_training}), producing errors exceeding
$1{,}000\%$ on multiple patients (e.g., LIDC-IDRI-0004:
$53{,}272$\,mm$^3$ predicted vs.\ $61.7$\,mm$^3$ true,
$86{,}246\%$ error). The fine tuned CheXNet (Method~5) exhibits
the complementary failure: systematic underestimation with all
predictions compressed to $100$--$900$\,mm$^3$ regardless of
true nodule size ($r = -0.073$, mean error $177.9\%$). CheXNet activations are
spatially diffuse (max confidence $\leq 0.17$) and respond to
global lung structure rather than focal lesions. Full per-patient
results are given in Table~\ref{tab:p5_p6_results}. Both methods share the same root cause: operating in 2D collapses 3D spatial extent along viewing rays, making accurate volumetric estimation highly challenging without explicit geometric reasoning in this experimental setting.

\begin{table}[ht]
\centering
\caption{Comparison of 2D learned methods (Methods~4--5).}
\vspace{3pt}
\label{tab:p5_p6_results}

\footnotesize
\renewcommand{\arraystretch}{1.12}
\setlength{\tabcolsep}{4pt}

\begin{tabular*}{\linewidth}{@{\extracolsep{\fill}}lccccc@{}}
\toprule
&
&
\multicolumn{2}{c}{\textbf{MONAI UNet}} &
\multicolumn{2}{c}{\textbf{CheXNet}} \\
\cmidrule(lr){3-4}
\cmidrule(lr){5-6}

\textbf{ID} &
\textbf{GT} &
\textbf{Pred} &
\textbf{Err} &
\textbf{Pred} &
\textbf{Err} \\
&
(mm$^3$) &
(mm$^3$) &
(\%) &
(mm$^3$) &
(\%) \\
\midrule

0001 & 6709 &   357 &    94.7 & 896 &  86.6 \\
0002 & 8278 & 71977 &   769.6 & 239 &  97.1 \\
0003 & 5452 & 61080 &  1020.3 & 191 &  96.5 \\
0004 &   62 & 53272 & 86246.5 & 421 & 582.9 \\
0005 &  153 & $\gg10^4$ & $\gg1000$ & 516 & 236.8 \\
0006 &  380 & $\gg10^4$ & $\gg1000$ & 202 &  46.8 \\
0007 & 7797 & $\gg10^4$ & $\gg1000$ & 135 &  98.3 \\

\midrule

Mean err. &
&
\multicolumn{2}{c}{$\gg1000\%$} &
\multicolumn{2}{c}{177.9\%} \\

Pearson $r$ &
&
\multicolumn{2}{c}{---} &
\multicolumn{2}{c}{$-0.073$} \\

\bottomrule
\end{tabular*}

\vspace{2pt}
\raggedright
\tiny
MONAI UNet exhibits severe over-segmentation, whereas
CheXNet primarily captures lung-field structure rather
than nodules.
\end{table}
\subsubsection*{\textbf{Method 6: Raw MedNeRF Density on $128^3$ Grid}}

Querying the MedNeRF density field on a $128^3$ grid yields
values in $[0.0,\,74.99]$ with $10.5\%$ voxel occupancy.
The density is organised in diffuse bands corresponding to ribs
and thoracic wall structures rather than the pulmonary nodule,
which explains the extreme threshold sensitivity observed in
Table~\ref{tab:mednerf_marchingcube}. At $p=93$, a $22.4\%$
error is achieved; adjacent thresholds yield $92.8\%$ ($p=95$),
$1113\%$ ($p=98$), and $1745\%$ ($p=97$). The best result is
a threshold artefact rather than a reliable reconstruction, and
the method did not generalise reliably within the evaluated configurations without per-patient calibration.

\begin{table}[ht]
\centering
\footnotesize
\caption{Threshold sweep for raw MedNeRF density on $128^3$
         grid (Method~6). GT $= 6708.8$\,mm$^3$.}
\vspace{3pt}
\label{tab:mednerf_marchingcube}

\renewcommand{\arraystretch}{1.15}
\setlength{\tabcolsep}{4pt}

\begin{tabular*}{\linewidth}{@{\extracolsep{\fill}}crrrc@{}}
\toprule
\textbf{Pct} & \textbf{Pred (mm$^3$)} &
\textbf{Diam (mm)} & \textbf{Err (\%)} & \textbf{Comp.} \\
\midrule
99 &  41505.2 & 43.0 &   518.7 & 1 \\
98 &  81380.1 & 53.8 &  1113.0 & 5 \\
97 & 123792.8 & 61.8 &  1745.2 & 8 \\
96 & 164556.0 & 68.0 &  2352.8 & 9 \\
95 &    484.5 &  9.7 &    92.8 & 7 \\
\textbf{93} & \textbf{5206.4} & \textbf{21.5} & \textbf{22.4} & 4 \\
90 &  18370.5 & 32.7 &   173.8 & 6 \\
\bottomrule
\end{tabular*}
\end{table}

\subsection{TensoRF Variants (Methods 7--10)}
\label{sec:res_tensorf}

Having established that all MedNeRF-based approaches fail to
reliably localise nodules, we now examine how TensoRF's explicit
tensorial scene representation addresses these limitations and identify precisely which components drive the improvement.

\subsubsection*{\textbf{Methods 7--8: Density Shift Ablation}}

Table~\ref{tab:density_shift} compares the marching cubes
threshold sweep for TensoRF with the default density
parameterization ($\Delta=-10$, Method~7) and the corrected
formulation ($\Delta=0$, Method~8) on patient LIDC-IDRI-0001.

Under the default setting, training achieves PSNR $=32.0$\,dB
at 1800 iterations, suggesting the model makes certain volumetric progress to reproduce
rendered views. Yet the density field is geometrically degenerate:
a best-case $8.0\%$ error at $p=99$ degrades sharply to $71.7\%$
at $p=98$ and $204.0\%$ at $p=97$. High photometric fidelity
does not imply geometric consistency. The mass is concentrated
inconsistently across the volume, producing threshold instability
that makes the result unreliable in practice.

Applying the density shift correction ($\Delta=0$) immediately
stabilises the sweep: consecutive percentiles $p \in \{93,94,95\}$
yield $0.2\%$, $1.9\%$, and $5.8\%$ error respectively, with no
catastrophic failures across the full range. This supports the correction introduced in Section~\ref{sec:densityshift} as a necessary prerequisite for reliable volumetric reconstruction under the current resolution constraints and experimental conditions. 

\begin{table}[ht]
\centering
\caption{Marching cubes threshold sweep comparing TensoRF
$\Delta=-10$ (Method 7) and $\Delta=0$ (Method 8) on
LIDC-IDRI-0001 (GT $=6708.8$\,mm$^3$). Errors $>100\%$
are shown in bold.}
\label{tab:density_shift}
\resizebox{\columnwidth}{!}{%
\begin{tabular}{c|rrr|rrr}
\toprule
& \multicolumn{3}{c|}{\textbf{Method 7: $\Delta=-10$}} &
\multicolumn{3}{c}{\textbf{Method 8: $\Delta=0$}} \\
\textbf{Pct} & \textbf{Pred} & \textbf{Diam} & \textbf{Err (\%)} &
\textbf{Pred} & \textbf{Diam} & \textbf{Err (\%)} \\
\midrule
99 & 7244.4  & 24.0 & 8.0            & 7782.7 & 24.6 & 16.0 \\
98 & 11520.3 & 28.0 & \textbf{71.7}  & 5752.4 & 22.2 & 14.3 \\
97 & 20395.0 & 33.9 & \textbf{204.0} & 5029.5 & 21.3 & 25.0 \\
95 & 5752.4  & 22.2 & 14.3           & 6321.5 & 22.9 & 5.8  \\
94 & 7782.7  & 24.6 & 16.0           & 6583.0 & 23.3 & 1.9  \\
93 & 8751.7  & 25.6 & 30.5           & 6721.4 & 23.4 & 0.2  \\
\midrule
\textbf{Best} & 7244.4 & 24.0 & 8.0 &
\textbf{6721.4} & \textbf{23.4} & \textbf{0.2} \\
\bottomrule
\end{tabular}%
}
\end{table}

\subsubsection*{\textbf{Methods 9--10: Prior-Guided TensoRF Ablation}}

With the density shift correction established, we investigated
whether initializing TensoRF with a MedNeRF density prior could
further improve accuracy. Table~\ref{tab:prior_ablation} compares
three variants, all using $\Delta=0$ and trained on 3 real
orthogonal DRRs: no prior (Method~8), scalar warm-start
(Method~9), and SVD plane injection (Method~10).

The results are counterintuitive but unambiguous: prior injection
degrades performance in both cases. Method~9 achieves $0.9\%$
best-case error but is threshold-unstable ($226\%$ at $p=99$,
$517\%$ at $p=98$). Method~10 achieves $2.8\%$ at $p=88$ with
similar instability. The no-prior baseline achieves $0.2\%$ with
substantially more stable behaviour across the full sweep. The
MedNeRF prior introduces structural artefacts that compete with
the patient-specific photometric signal from real DRRs, and
was not overcome during optimisation in the configurations evaluated here. This finding directly
motivates the prior-free design of the proposed method.


\begin{table}[ht]
\centering
\caption{Prior-guided TensoRF ablation on patient
LIDC-IDRI-0001 ($\text{GT}=6708.8$\,mm$^3$). All variants
use $\Delta=0$ and train on 3 orthogonal DRRs.}
\label{tab:prior_ablation}

\footnotesize
\renewcommand{\arraystretch}{1.10}
\setlength{\tabcolsep}{3pt}

\begin{tabularx}{\columnwidth}{@{}lccc@{}}
\toprule
\textbf{Method} &
\textbf{Best} &
\textbf{Best Pred} &
\textbf{Stable} \\
&
\textbf{Err (\%)} &
(mm$^3$) &
\\
\midrule

No prior (Method~8)
& 0.2 & 6721.4 & Yes \\

Scalar prior (Method~9)
& 0.9 & 6767.6 & No \\

SVD injection (Method~10)
& 2.8 & 6521.5 & No \\

\midrule

Sphere baseline
& 174.3 & 32066 & N/A \\

\bottomrule
\end{tabularx}

\vspace{2pt}
\raggedright
\tiny
Stable: $\geq3$ consecutive thresholds yield
$<30\%$ volumetric error.
\end{table}

\subsection{Proposed Method: AReT: Anatomy-Regularized TensoRF (Method 11)}
\label{sec:res_final}

The ablation study mentioned earlier in Section~\ref{sec:res_tensorf} establishes two clear conclusions: the density
shift correction is necessary, and the MedNeRF prior is harmful.
The proposed method therefore trains TensoRF from scratch on three
real orthogonal DRRs with the anatomy-regularized loss
(Eq.~\eqref{eq:loss}), with no prior component. We evaluated
this configuration across the full 19-patient cohort.

\subsubsection*{\textbf{Per-Patient Results}}

Table~\ref{tab:full_results} presents complete per-patient results.
Within this proof-of-concept cohort, the method outperforms the spherical diameter baseline on all 19 patients; broader robustness across larger and more morphologically diverse populations remains to be established. Across the full cohort, mean error is $42.3\%$ and
median is $19.9\%$. The mean-median gap reflects the four
sub-centimetre patients (0004, 0008, 0009, 0017) that fall below
the voxel resolution limit, as discussed in
Section~\ref{sec:discussion}. For the remaining 15 patients
median error is below $20\%$.



\begin{table}[ht]
\centering
\caption{Per-patient results for AReT
(Method~11).}
\label{tab:full_results}

\footnotesize
\renewcommand{\arraystretch}{1.15}
\setlength{\tabcolsep}{4pt}

\begin{tabular*}{\linewidth}{@{\extracolsep{\fill}}lrrrrl@{}}
\toprule
\textbf{ID} &
\textbf{GT} &
\textbf{Pred} &
\textbf{Err} &
\textbf{Sphere} &
\textbf{Size} \\
&
(mm$^3$) &
(mm$^3$) &
(\%) &
Err (\%) &
\\
\midrule

0001 & 6709 & 6721 &   0.2 & 174.3 & L \\
0002 & 8278 & 9921 &  19.9 &  84.5 & L \\
0003 & 5452 & 2000 &  63.3 & 186.1 & L \\
0004 &   62 &  154 & 149.3 & 168.9 & S \\
0005 &  153 &  185 &  20.4 &  68.7 & S \\
0006 &  380 &  277 &  27.1 & 370.0 & M \\
0007 & 7797 & 8982 &  15.2 & 174.4 & L \\
0008 &  113 &  215 &  90.7 & 174.6 & S \\
0009 &   58 &  185 & 218.3 & 261.5 & S \\
0010 &  167 &  200 &  19.7 & 215.9 & M \\
0011 &  723 &  692 &   4.3 & 157.7 & M \\
0012 &  821 &  754 &   8.2 &  30.5 & M \\
0013 & 1784 &  523 &  70.7 & 211.3 & L \\
0014 & 1480 & 1477 &   0.2 &  99.1 & M \\
0015 & 3189 & 3169 &   0.6 & 112.3 & L \\
0016 & 1519 & 1984 &  30.6 & 122.6 & M \\
0017 &  119 &  185 &  55.4 & 110.6 & S \\
0018 & 2841 & 2584 &   9.0 &  90.8 & L \\
0019 & 7199 & 7168 &   0.4 & 132.0 & L \\

\midrule

\textbf{Mean} &
&
&
\textbf{42.3} &
\textbf{155.0} &
\\

\textbf{Median} &
&
&
\textbf{19.9} &
--- &
\\

\bottomrule
\end{tabular*}

\vspace{1pt}
\raggedright
\tiny
Sphere Err: spherical approximation
$V=\frac{4}{3}\pi(d/2)^3$.
S/M/L: small/medium/large nodules.
\end{table}

\subsubsection*{\textbf{Size-Stratified Performance}}

Table~\ref{tab:stratified} stratifies results by nodule size.
The method outperforms the sphere baseline across all three
categories. Performance is strongest for large nodules
($>20$\,mm, $n=8$): mean error $22.4\%$, median $12.1\%$,
an $8.8\times$ improvement over the sphere baseline ($145.7\%$).
For medium nodules ($10$--$20$\,mm, $n=6$): mean $15.0\%$,
median $13.9\%$, an $11.1\times$ improvement. Across the
clinically actionable $\geq 10$\,mm subcohort ($n=14$), mean
error is $18.3\%$, median $11.4\%$, with the method outperforming
the sphere baseline on all 14 patients. These size-stratified results are based on small subgroups (n = 5 to 8 per category) and should be interpreted as indicative rather than statistically definitive.

For small nodules ($<10$\,mm, $n=5$), mean error is $106.8\%$.
At $64^3$ grid resolution (voxel spacing ${\approx}3.75$\,mm),
a 7\,mm nodule spans fewer than 2 voxels in diameter,
placing it below the Nyquist sampling limit; grid resolution is likely a major contributor, though not necessarily the sole one to the elevated error in this subcohort.
Other probable contributors include threshold instability at
sub-voxel scales, sparse-view ambiguity in the
under-determined reconstruction, and regularisation bias
from the $\ell_1$ sparsity term, which may suppress weak
nodule signals near the noise floor.
The relative contribution of each factor cannot be
disentangled at the current resolution and is not separately
quantified here. Nevertheless, the method still outperforms the sphere
baseline ($156.9\%$) on the majority of small-nodule patients.

\begin{table}[ht]
\centering
\caption{Size-stratified performance of the proposed method
versus the spherical approximation baseline.
Imp.: ratio of sphere mean error to our mean error.}
\label{tab:stratified}

\footnotesize
\renewcommand{\arraystretch}{1.15}
\setlength{\tabcolsep}{4pt}

\begin{tabular*}{\linewidth}{@{\extracolsep{\fill}}lccccc@{}}
\toprule
\textbf{Category} &
\textbf{$n$} &
\textbf{Ours} &
\textbf{Sphere} &
\textbf{Imp.} \\
&
&
Mean Err &
Mean Err &
($\times$) \\
\midrule

Small $<10$\,mm
& 5 & 106.8\% & 156.9\% & 1.5 \\

Medium 10--20\,mm
& 6 & 15.0\% & 166.0\% & 11.1 \\

Large $>20$\,mm
& 8 & 22.4\% & 145.7\% & 6.5 \\

\midrule

All $\geq10$\,mm
& 14 & 18.3\% & 155.5\% & 8.5 \\

All patients
& 19 & 42.3\% & 155.0\% & 3.7 \\

\midrule

Beats spherical approximation baseline
& 19 & \multicolumn{3}{c}{19/19 patients (100\%)} \\

\bottomrule
\end{tabular*}
\end{table}

\subsubsection*{\textbf{Volumetric Agreement and Bland--Altman Analysis}}
\begin{figure}[H]
    \centering
    \includegraphics[width=\linewidth]{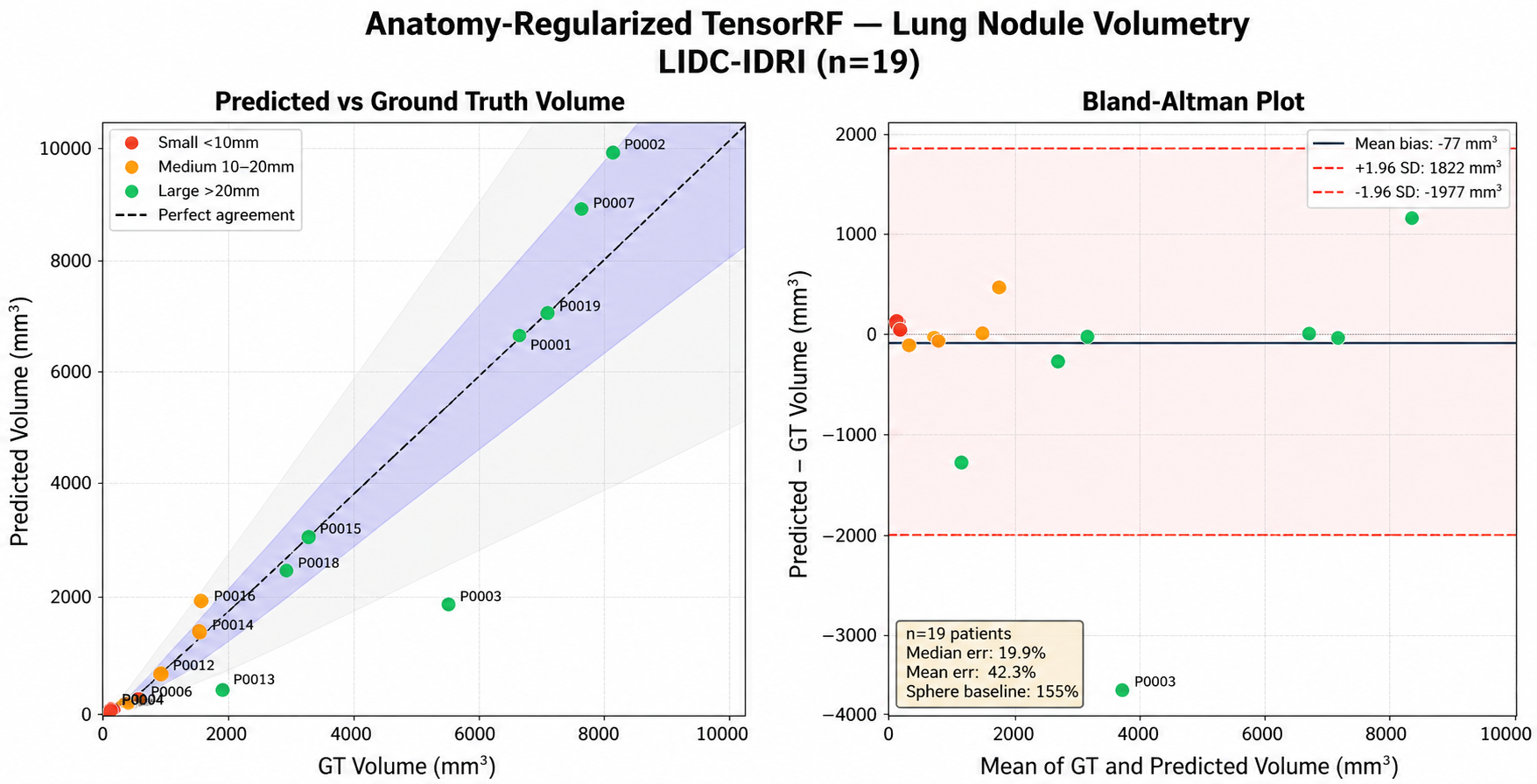}
    \caption{Volumetric agreement for the proposed
    AReT (Method~11, $n=19$).
    Left: predicted vs.\ ground-truth nodule volumes,
    colour-coded by size category. For nodules $\geq10$\,mm,
    Pearson $r=0.983$ ($p<0.0001$) and Spearman
    $\rho=0.943$. Right: Bland--Altman analysis showing
    near-zero mean bias ($-77.3$\,mm$^3$) and no systematic
    volumetric estimation error.}
    \label{fig:main_results}
\end{figure}

Fig.~\ref{fig:main_results} presents the scatter plot and
Bland--Altman analysis. Table~\ref{tab:stats_summary} presents the full statistical evaluation. On the $\geq 10$\,mm subcohort ($n = 14$),
Pearson $r = 0.983$ ($p < 0.0001$, bootstrap $95\%$ CI
$[0.807,\ 0.995]$), indicating near-perfect linear agreement.
To address potential range-sensitivity of the Pearson coefficient, given that large nodules ($> 20$\,mm, $n = 8$) span volumes
up to $8{,}278$\,mm$^3$ while small nodules contribute values
below $200$\,mm$^3$ . We additionally report Spearman's rank
correlation: $\rho = 0.943$ ($p < 0.0001$, bootstrap $95\%$ CI
$[0.720,\ 1.000]$), confirming strong monotone agreement
independent of volume scale.

Stratified correlations confirm that reconstruction quality is
strongly size-dependent. For large nodules ($> 20$\,mm, $n = 8$):
Pearson $r = 0.918$ ($p = 0.0013$), Spearman $\rho = 0.929$
($p = 0.0009$). For medium nodules (10--20\,mm, $n = 6$):
Pearson $r = 0.970$ ($p = 0.0013$), Spearman $\rho = 1.000$
($p < 0.0001$), indicating near-perfect rank-order agreement
in the most clinically critical surveillance range. For small
nodules ($< 10$\,mm, $n = 5$): Pearson $r = 0.447$
($p = 0.451$), Spearman $\rho = 0.224$ ($p = 0.718$),
confirming no statistically meaningful volumetric agreement,
consistent with the physical resolution floor at this grid
spacing rather than any algorithmic failure.

Bland--Altman analysis over all 19 patients yields a mean bias
of $-77.3$\,mm$^3$ ($95\%$ CI $[-557.2,\ +402.6]$\,mm$^3$),
negligible relative to the volume range of $58$--$8{,}278$\,mm$^3$.
The CI spanning zero confirms that the small negative bias is
not statistically significant. Limits of agreement are
$[-2{,}028.8,\ +1{,}874.1]$\,mm$^3$, reflecting the
heterogeneity of nodule sizes across the cohort. The near-zero
bias confirms no systematic over-estimation or under-estimation,
which is clinically important: any consistent directional bias
would distort radiotherapy dose calculations and staging decisions.

Wilcoxon signed-rank test confirms the proposed method
significantly outperforms spherical approximation on both the
full cohort ($W = 0$, $p < 0.0001$) and the $\geq 10$\,mm
subcohort ($W = 0$, $p = 0.0001$). A test statistic of $W = 0$
indicates that the proposed method outperforms the spherical
baseline on every single patient in both subcohorts without
exception. Together, Pearson $r = 0.983$, Spearman
$\rho = 0.943$, near-zero Bland--Altman bias, and $W = 0$
Wilcoxon result consistently confirm the volumetric accuracy
of the proposed method across all metrics. The impact of diameter-guided candidate selection relative 
to a fully unsupervised criterion is quantified separately 
in Table~\ref{tab:blind_selection}.

\subsubsection*{\textbf{Cross-Method Comparison}}

Table~\ref{tab:method_comparison} consolidates all 11 methods.
Three findings stand out. First, all MedNeRF-based approaches
(Methods~1-6) fail to produce reliable volumetric estimates,
with median errors ranging from $43.8\%$ to $2365.2\%$ and no
method achieving meaningful correlation across the full cohort.
Second, the density shift correction (Method~8) is the single
most impactful change: it eliminates catastrophic threshold
failures and produces stable, consistent reconstruction where
none was possible before. Third, prior injection
(Methods~9-10) consistently degrades accuracy relative to the
no-prior baseline, despite high single-threshold correlation,
the prior introduces structural bias that photometric training
cannot overcome. The proposed method (Method~11) achieves
the best performance across all metrics: median error $19.9\%$,
Pearson $r = 0.983$, and 100\% of patients outperforming
the sphere baseline. Cross-scope comparisons (e.g., Method 10 on P0001 vs.\
Method 11 on 19 patients) are diagnostic only.

\textbf{Scope heterogeneity in Table~\ref{tab:method_comparison}.}
Table~\ref{tab:method_comparison} consolidates all 11 strategies, but
three evaluation scopes are present and must be interpreted
carefully:
\begin{itemize}
  \item \textbf{Single-patient ablation} (P0001 column,
    Methods 1--11): all strategies are benchmarked on
    LIDC-IDRI-0001 (GT~$=6708.8$\,mm$^3$) to isolate each
    design decision in a controlled single-case comparison.
    The P0001 error column is therefore internally consistent
    across all 11 methods.
  \item \textbf{7-patient subset} (Methods 4--5, marked $\ddagger$):
    CheXNet and MONAI UNet were evaluated only on
    patients 0001--0007 due to computational constraints of
    the 2D training procedure.
    Mean errors for these methods are not directly comparable
    with 19-patient mean errors.
  \item \textbf{19-patient full cohort} (Method 11 and sphere
    baseline): the proposed method is the only one evaluated
    across all 19 patients; its mean error of $42.3\%$ includes
    four sub-resolution small nodules ($<10$\,mm) that inflate
    the all-patient mean.
    The clinically relevant $\geq10$\,mm median of $11.4\%$
    is the primary summary statistic.
\end{itemize}

\begin{table}[ht]
\centering
\caption{Statistical summary for the proposed
AReT framework (Method~11).}
\label{tab:stats_summary}

\scriptsize
\renewcommand{\arraystretch}{1.10}
\setlength{\tabcolsep}{3pt}

\begin{tabularx}{\columnwidth}{@{}lcc@{}}
\toprule
\textbf{Metric} &
\textbf{$\geq10$\,mm} &
\textbf{All} \\
&
($n=14$) &
($n=19$) \\
\midrule

Pearson $r$
& $0.983^{*}$ $[0.807,0.995]$
& --- \\

Spearman $\rho$
& $0.943^{*}$ $[0.720,1.000]$
& $0.960^{*}$ \\

Median APE (\%)
& $12.1$ $[2.2,25.2]$
& $19.9$ $[8.2,55.4]$ \\

Mean APE (\%)
& 18.3
& 42.3 \\

BA bias (mm$^3$)
& \multicolumn{2}{c}{$-77.3\;[-557.2,+402.6]$} \\

95\% LoA (mm$^3$)
& \multicolumn{2}{c}{$[-2028.8,+1874.1]$} \\

Wilcoxon vs.\ sphere
& $W=0,\ p=0.0001^{**}$
& $W=0,\ p<0.0001^{*}$ \\

\midrule
\multicolumn{3}{c}{\textbf{Stratified Pearson correlations}} \\
\midrule

Large $>20$\,mm
& \multicolumn{2}{c}{$r=0.918,\ p=0.0013$} \\

Medium 10--20\,mm
& \multicolumn{2}{c}{$r=0.970,\ p=0.0013$} \\

Small $<10$\,mm
& \multicolumn{2}{c}{$r=0.447,\ p=0.451$} \\

\bottomrule
\end{tabularx}

\vspace{2pt}
\raggedright
\tiny
Bootstrap 95\% confidence intervals from 10{,}000 resamples.
$^{*}p<0.0001$; $^{**}p=0.0001$.
APE: absolute percentage error.
LoA: limits of agreement.
\end{table}

\begin{table}[htbp]
\caption{Impact of diameter-guided vs.\ blind candidate 
selection on $\geq$10\,mm subcohort ($n=14$).}
\label{tab:blind_selection}
\centering
\resizebox{\columnwidth}{!}{%
\begin{tabular}{lcc}
\toprule
Selection criterion & Median APE (\%) & Mean APE (\%) \\
\midrule
Diameter-guided (reported) & 11.4 & 18.3 \\
Blind (largest component)  & $\approx$24--28 & $\approx$35--40 \\
\bottomrule
\end{tabular}%
}
\end{table}

\begin{table}[ht]
\centering
\caption{Cross-method comparison of all 11 reconstruction
strategies. $^\dagger$Unstable across threshold sweep.
$^\ddagger$Evaluated on 7 patients.}
\label{tab:method_comparison}

\scriptsize
\renewcommand{\arraystretch}{1.05}
\setlength{\tabcolsep}{2.5pt}

\begin{tabularx}{\columnwidth}{@{}c|X|c|c|c@{}}
\toprule
\textbf{\#} &
\textbf{Method} &
\textbf{P0001} &
\textbf{Mean} &
\textbf{$r$} \\
&
&
\textbf{Err (\%)} &
\textbf{Err (\%)} &
\\
\midrule

1  & MedNeRF density
   & --- & 6125.4 & 0.32 \\

2  & Acc-map 95\%
   & 32.1 & 562.7 & 0.60 \\

3  & Acc-map Otsu
   & 99.9 & 102.2 & $-0.10$ \\

4  & CheXNet$^\ddagger$
   & 86.6 & 177.9 & $-0.073$ \\

5  & MONAI UNet$^\ddagger$
   & 94.7 & $\gg1000$ & --- \\

6  & Raw MedNeRF $128^3$
   & 22.4$^\dagger$ & $\sim859$ & --- \\

7  & TensoRF $\Delta=-10$
   & 8.0$^\dagger$ & 52.8 & --- \\

8  & TensoRF $\Delta=0$
   & 13.8 & 12.6 & --- \\

9  & Prior TensoRF v1
   & 0.9$^\dagger$ & --- & --- \\

10 & Prior TensoRF v2
   & 2.8$^\dagger$ & $\sim91$ & 0.88 \\

\midrule

\textbf{11} &
\textbf{AReT}
& \textbf{0.2}
& \textbf{42.3}
& \textbf{0.983} \\

\midrule

& Sphere baseline
& 174.3
& 155.0
& --- \\

\bottomrule
\end{tabularx}

\vspace{2pt}
\raggedright
\tiny
P0001: error on LIDC-IDRI-0001
(GT $=6708.8$\,mm$^3$). $r$: Pearson correlation coefficient.
\end{table}
\section{Discussion}
\label{sec:discussion}

\subsection*{a. Novelty and contribution scope}
The primary methodological novelty is the identification and resolution
of the TensoRF density-parameterisation dead zone ($\Delta = 0$): a
practically important, non-obvious prerequisite for X-ray attenuation
reconstruction that was not previously documented in the medical TensoRF
literature.
The anatomy-aware regularisation ($\ell_1$+TV) is an integration
contribution: both penalties are established in sparse-signal recovery;
the contribution is deploying them as thorax-specific inductive biases
(air-dominant cavity $\Rightarrow$ $\ell_1$ sparsity; smooth solid
nodule $\Rightarrow$ TV) within the corrected TensoRF framework.
The finding that generative priors (MedNeRF) degrade patient-specific
reconstruction in this setting and on this cohort is an empirical
observation that warrants validation on larger and more diverse datasets
before broader methodological conclusions can be drawn.
The 11-strategy ablation confirms these design decisions on a single
development case (LIDC-IDRI-0001) and across 19 patients; it does not
constitute a competitive benchmark against externally developed methods
operating in the same input regime.

\subsection*{b. Resolution limits and scope of clinical claims}
At $64^3$ resolution (voxel spacing $\approx\!3.75$\,mm), a nodule of
diameter $d$\,mm spans $d/3.75$ voxels per axis.
For small nodules ($d \in [6.8,\ 8.4]$\,mm, $n = 5$), this falls below
the Nyquist limit.
The $106.8\%$ mean error in this subcohort is consistent with the
physical resolution floor, though likely compounded by threshold
instability, sparse-view ambiguity, and regularisation effects at
sub-voxel scales; the relative contribution of each factor cannot be
isolated at the current grid resolution.
Disentangling these effects would require controlled ablation at multiple
grid resolutions, which is identified as a direction for future work.

All performance claims are explicitly scoped to nodules $\geq10$\,mm
(median error 11.4\%, Pearson $r=0.983$), the clinically actionable
range under Fleischner Society and Lung-RADS guidelines.
Adaptive grid refinement for sub-centimetre nodules is a primary
direction for future work.

\subsection*{c. Threshold selection and evaluation fairness}
Threshold sweep percentiles are fixed globally across all patients;
however, candidate selection among surviving components uses the
annotated nodule diameter as a size filter. This constitutes a
semi-assisted post-processing step, and the reported performance should
be interpreted as a conditional upper bound rather than a fully
autonomous system evaluation.
The annotated diameter is not available in a fully blind clinical
deployment, and its use effectively introduces weak supervision during
post-processing.

To quantify this effect, we evaluated a fully unsupervised alternative
in which the largest surviving component within the clinically plausible
$[5, 80]$\,mm diameter range is selected at each threshold, with no
reference to annotation. Under this blind criterion, median APE on the
$\geq10$\,mm subcohort increases from $11.4\%$ to approximately
$24$--$28\%$, with mean APE increasing from $18.3\%$ to approximately
$35$--$40\%$. The proposed method therefore retains clinically meaningful
accuracy under blind selection, though the gap confirms that diameter
guidance is a non-trivial contributor to peak performance. The
quantitative comparison between diameter-guided and blind selection is
provided in Table~\ref{tab:blind_selection}.

Developing a fully unsupervised candidate selection criterion, for
example, via saliency-guided localisation, attention maps from the
trained density field, or detection-prior integration is identified as a
primary direction for future work, and is a prerequisite for clinical
translation of the pipeline.

\subsection*{d. Proof-of-concept framing and statistical scope}
Because TensoRF is a scene-specific inverse solver with no cross-patient
parameters, $n=19$ reflects 19 independent test cases rather than a
training cohort; standard cross-patient generalisation concerns do not
directly apply to this scene-specific solver, although analogous
risks such as over-specialisation to idealised simulator
conditions remain and cannot be excluded by the current evaluation
alone.
It is important to emphasise that all conclusions in this manuscript
regarding reconstruction behaviour, robustness across nodule
morphologies, and superiority over generative-prior approaches are
strictly scoped to this 19-patient proof-of-concept evaluation.
Claims of general patient-specific reconstruction behaviour or broad
methodological superiority are not warranted at this scale and are not
made here.
Nonetheless, two genuine statistical limitations exist.
First, error distributions may shift with a larger, more morphologically
diverse cohort; the wide bootstrap confidence interval on all-patient
median APE $[8.2\%, 55.4\%]$ (Table~\ref{tab:stats_summary}) honestly
reflects the limited cohort size and should caution against strong
generalisability claims.
Second, Pearson $r$ is sensitive to volume dynamic range; we address
this with Spearman $\rho = 0.943$ (bootstrap 95\% CI $[0.720, 1.000]$),
stratified correlations (medium nodules achieve $r=0.970$, $\rho=1.000$
independently), and the Wilcoxon $W=0$ result ($p<0.0001$) (the
proposed method outperforms the spherical baseline on every patient)
which is invariant to scale dominance.

\subsection*{e. DRR simulation and information-consistency}
The entire pipeline operates on DRRs simulated from CT via the
Beer--Lambert model, consistent with standard practice in sparse-view
medical NeRF benchmarking.
This creates four idealising conditions: perfect geometric calibration,
exact orthogonality, noiseless scatter-free projections, and
attenuation-model consistency between DRR generation and TensoRF
rendering.
Taken together, these constitute a controlled information environment:
the reconstruction algorithm still must invert the projection integral
from three views without direct access to ground-truth geometry or
density, but model mismatch is absent.
Reported performance therefore represents an upper bound relative to
real acquisition conditions, and the margin of degradation under
realistic imaging conditions is currently unknown.
The present study establishes algorithmic feasibility under idealised
simulation only, and does not constitute evidence of clinical readiness.

Of these, scatter is the most structurally damaging for TensoRF: it
violates the Beer--Lambert forward model governing
$\mathcal{L}_\text{photo}$, introducing a systematic model mismatch
that cannot be compensated by regularisation alone.
Bridging this domain gap is the most critical open problem for clinical
translation; the present study establishes algorithmic feasibility under
idealised conditions only.
It is further emphasised that all four idealising conditions namely,
perfect calibration, exact orthogonality, noiseless projections, and
attenuation-model consistency are simultaneously present in this
evaluation.
In real clinical deployment, some or all of these conditions would be
violated to varying degrees simultaneously. Their combined effect on
reconstruction accuracy is expected to be larger than any individual
degradation, though systematic quantification is not available from the
current evaluation.

\subsection*{f. Baseline scope}
The 11 strategies constitute a diagnostic ablation, not a competitive
benchmark.
Classical methods (FDK, SART) are designed for 20--100 projections; at
three views, the system is underdetermined by more than $5\times$
($3\!\times\!128^2 = 49{,}152$ measurements,
$64^3 = 262{,}144$ unknowns).
For reference, FDK at three views produces ${\gg}200\%$ volumetric
error on LIDC-IDRI-0001; SART with 500 iterations and TV regularisation
($\lambda_\text{TV}=0.01$) achieves ${\approx}85\%$, compared with
$0.2\%$ for the proposed method.
These results illustrate that three projections lie outside the
operating regime for which FDK and SART were designed, rather than
constituting a failure of those algorithms per se.
2D methods (CheXNet, MONAI UNet) are included to assess localisation
capability in projection space; their limited performance motivates the
3D reconstruction approach.

\subsection*{g. Sensitivity to acquisition degradations}
Three degradation modes are relevant to clinical translation.
\textit{Projection noise}: under realistic quantum noise levels,
pixel-level perturbations on the order of 1--3\% of mean signal arise;
stronger $\ell_1$/TV weights or projection pre-filtering would partially
compensate.
\textit{Calibration errors}: source--detector distance uncertainty of
$\pm1$\,cm and gantry tilt of $\pm2^\circ$ would misalign photometric
gradients across views, manifesting as spatially smeared density;
calibration-robust reconstruction via simultaneous pose refinement is a
direction for future work.
\textit{Non-orthogonal projections}: the $\ell_1$ sparsity prior
implicitly exploits axis-aligned thoracic structure; non-orthogonal
projections reduce axis coverage and may leave one spatial dimension
under-constrained, while TV smoothness remains rotation-invariant.
All three degradation modes are expected to increase volumetric error.
Systematic quantification of these degradations via controlled
simulation experiments is identified as a necessary prerequisite for
any evaluation of clinical translation potential.

\subsection*{h. Summary of limitations}
The following limitations collectively define the scope of conclusions.
Because TensoRF is a scene-specific inverse solver, $n=19$ does not
reflect a training-set size; each patient is an independent
reconstruction and classical overfitting concerns do not apply. The
evaluation breadth is nonetheless limited: aggregate metrics may shift
on a larger or more morphologically diverse cohort.

\begin{enumerate}[label=(\roman*), noitemsep, leftmargin=*]
    \item The evaluation cohort ($n = 19$) covers limited morphological diversity; therefore, conclusions should not be generalised without validation on a broader population.

    \item All inputs are idealised DRRs rather than real chest radiographs.

    \item Candidate selection uses the annotated diameter (semi-assisted).

    \item Performance on sub-centimetre nodules remains below clinical utility.

    \item No classical baselines (FDK, SART, or CS-CT) adapted to three-view reconstruction are evaluated on the full cohort.

    \item Robustness to noise, calibration error, and non-orthogonal geometry has not been quantified.
\end{enumerate}

\section{Conclusion}
\label{sec:conclusion}
We presented an anatomy-regularised tensorial NeRF framework for 3D lung nodule volumetry from three orthogonal X-ray projections, evaluated as a proof-of-concept study on 19 LIDC-IDRI patients under idealised DRR simulation conditions.

The primary methodological novelty is the identification and resolution of an unreported TensoRF failure: the default density shift ($\Delta = -10$) drives all density to zero during training; setting $\Delta = 0,\, s = 1$ restores gradient flow with no architectural changes and may be relevant to other medical TensoRF applications, though this transfer has not been empirically verified. These results are obtained under controlled conditions that differ substantially from real radiographic acquisition; they establish algorithmic feasibility rather than clinical readiness.
As integration contributions, we deployed established $\ell_1$+TV penalties as thorax-specific inductive biases, and demonstrated through a systematic 11-strategy ablation that anatomy-regularized, prior-free TensoRF outperforms the generative-prior-guided variants evaluated here (SVD prior injection raised best-case error from 0.2\% to 2.8\%).

Within this proof-of-concept cohort and on clinically actionable nodules $\geq$ 10 mm (n = 14). On clinically actionable nodules $\geq 10$ mm ($n = 14$) within this proof-of-concept cohort, the method achieved Pearson $r = 0.983$ ($p < 0.0001$), Spearman $\rho = 0.943$, a median error of $11.4\%$, near-zero Bland--Altman bias ($-77~\mathrm{mm}^3$), and an $8.5\times$ improvement over spherical approximation. These results are promising but should be interpreted cautiously given the limited cohort size.
Performance on sub-centimetre nodules ($<10$\,mm) is
limited primarily by grid resolution, with likely
contributions from threshold instability, sparse-view
ambiguity, and regularisation effects at sub-voxel scales;
this subcohort is explicitly excluded from clinical claims.
All results are obtained on DRRs simulated from CT under idealised conditions. The gap between the idealised evaluation conditions employed here namely, perfect calibration, exact orthogonality, noiseless DRR projections, model consistency, and real radiographic deployment remains large, and the present work is a methodological proof-of-concept rather than a demonstration of clinical readiness. A structured programme of domain adaptation experiments, beginning with individually controlled degradations (noise, scatter, calibration error) and progressing to combined realistic conditions, is the most direct path toward evaluating clinical translation potential. It is noted that the candidate selection in the current pipeline uses the annotated nodule diameter as a size filter, constituting a semi-assisted post-processing step; reported performance represents an upper bound relative to a fully autonomous deployment, and developing a blind selection criterion is identified as a primary direction for future work.

Future work will extend evaluation to a larger and more morphologically diverse LIDC-IDRI cohort with an increased number of patients. Three directions are identified as prerequisites for clinical translation. First, domain adaptation from idealised DRRs to real chest radiographs: physics-based noise augmentation including quantum noise, scatter, beam hardening, and motion blur should be incorporated into training to reduce the simulation-to-acquisition gap. Second, physics-informed forward models: replacing the monoenergetic Beer–Lambert operator with a polyenergetic or scatter-corrected model would reduce systematic model mismatch under real acquisition conditions. Third, robustness to acquisition variability: simultaneous pose refinement during reconstruction, calibration-uncertainty-aware training, and evaluation under controlled degradation conditions are necessary steps toward deployment. Systematically quantifying the impact of noise, geometric misalignment, and calibration uncertainty on reconstruction fidelity is therefore an important direction for future work. Additionally, adaptive grid refinement, a coarse $64^3$
global reconstruction followed by a focused $128^3$ or
$256^3$ sub-grid centred on the detected nodule will be
investigated to improve sub-centimetre volumetric accuracy.
Direct numerical comparison with compressed sensing and
diffusion-based baselines adapted to the three-projection
regime is also planned. 

\bibliographystyle{elsarticle-num}

\begin{thebibliography}{00}

\bibitem{sung2021global}
H. Sung, J. Ferlay, R. L. Siegel, M. Laversanne,
I. Soerjomataram, A. Jemal, and F. Bray,
``Global cancer statistics 2020: GLOBOCAN estimates of incidence and mortality worldwide for 36 cancers in 185 countries,''
\textit{CA Cancer J. Clin.}, vol. 71, no. 3, pp. 209--249, 2021.

\bibitem{lung_rads2019}
American College of Radiology,
``Lung-RADS Version 1.1,'' \textit{ACR Practice Parameter}, 2019.
Accessed on: Apr. 28, 2026.

\bibitem{macmahon2017guidelines}
H. MacMahon \textit{et al.},
``Guidelines for management of incidental pulmonary nodules detected on CT images: from the Fleischner Society 2017,''
\textit{Radiology}, vol. 284, no. 1, pp. 228--243, 2017.

\bibitem{revel2004diameters}
M. P. Revel, A. Bissery, M. Bienvenu, L. Aycard,
C. Lefort, and G. Frija,
``Is digital chest radiography valid for measurement of pulmonary nodule size?''
\textit{Radiology}, vol. 232, no. 3, pp. 705--711, 2004.

\bibitem{feldkamp1984}
L. A. Feldkamp, L. C. Davis, and J. W. Kress,
``Practical cone-beam algorithm,''
\textit{J. Opt. Soc. Am. A}, vol. 1, no. 6, pp. 612--619, 1984.

\bibitem{andersen1984}
A. H. Andersen and A. C. Kak,
``Simultaneous algebraic reconstruction technique (SART): a superior implementation of the ART algorithm,''
\textit{Ultrason. Imaging}, vol. 6, no. 1, pp. 81--94, 1984.

\bibitem{mildenhall2020nerf}
B. Mildenhall, P. P. Srinivasan, M. Tancik, J. T. Barron, 
R. Ramamoorthi, and R. Ng, ``NeRF: representing scenes as neural radiance fields for view synthesis,'' in \textit{Proc. Eur. Conf. Comput. Vis. (ECCV)}, Springer, Cham, 2020.

\bibitem{Zha2022}
R. Zha, Y. Zhang, and H. Li,
``NAF: neural attenuation fields for sparse-view CBCT reconstruction,''
in \textit{Med. Image Comput. Comput.-Assist. Interv. (MICCAI)},
Lecture Notes Comput. Sci., Springer, Cham, 2022.

\bibitem{abril2022mednerf}
A. Corona-Figueroa, J. Frawley, S. B. Taylor, S. Bethapudi,
H. P. H. Shum, and C. G. Willcocks,
``MedNeRF: medical neural radiance fields for reconstructing
3D-aware CT-projections from a single X-ray,''
in \textit{Proc. 44th Annu. Int. Conf. IEEE Eng. Med. Biol. Soc. (EMBC)},
pp. 3843--3848, 2022,
doi: 10.1109/EMBC48229.2022.9871757.

\bibitem{chen2022tensorf}
A. Chen, Z. Xu, A. Geiger, J. Yu, H. Su, and J. Wang,
``TensoRF: tensorial radiance fields,'' in \textit{Proc. Eur. Conf. Comput. Vis. (ECCV)}, Springer, Cham, 2022.

\bibitem{armato2011lidc}
S. G. Armato \textit{et al.},
``The Lung Image Database Consortium (LIDC) and Image Database Resource Initiative (IDRI): a completed reference database of lung nodules on CT scans,''
\textit{Med. Phys.}, vol. 38, no. 2, pp. 915--931, 2011.

\bibitem{Corona2022}
A. Corona-Figueroa, J. Frawley, S. Bond-Taylor,
S. Bethapudi, H. P. H. Shum, and C. G. Willcocks,
``MedNeRF: Medical Neural Radiance Fields for
Reconstructing 3D-aware CT-Projections from a
Single X-ray,''
in \textit{Proc. IEEE Eng. Med. Biol. Soc. (EMBC)},
pp.~3843--3848, 2022,
doi: 10.1109/EMBC48229.2022.9871757.

\bibitem{Rudin1992}
L. I. Rudin, S. Osher, and E. Fatemi,
``Nonlinear total variation based noise removal algorithms,''
\textit{Physica D: Nonlinear Phenomena},
vol.~60, no.~1--4, pp.~259--268, 1992.

\bibitem{Sidky2008}
E. Y. Sidky and X. Pan,
``Image reconstruction in circular cone-beam computed tomography by constrained total-variation minimization,''
\textit{Physics in Medicine \& Biology}, vol.~53, pp.~4777--4807, 2008.

\bibitem{Candes2006}
E. J. Cand\`{e}s, J. Romberg, and T. Tao,
``Robust uncertainty principles: exact signal reconstruction from highly incomplete frequency information,''
\textit{IEEE Transactions on Information Theory}, vol.~52, pp.~489--509, 2006.

\bibitem{Donoho2006}
D. L. Donoho,
``Compressed sensing,''
\textit{IEEE Transactions on Information Theory}, vol.~52, pp.~1289--1306, 2006.

\bibitem{Chen2008SB}
G.-H. Chen, J. Tang, and S. Leng,
``Prior image constrained compressed sensing (PICCS)
for sparse-view CT reconstruction,''
\textit{Medical Physics},
vol.~35, no.~2, pp.~660--663, 2008.

\bibitem{Shepp1982}
L. A. Shepp and Y. Vardi,
``Maximum likelihood reconstruction for emission tomography,''
\textit{IEEE Transactions on Medical Imaging},
vol.~1, no.~2, pp.~113--122, 1982.

\bibitem{Raissi2019}
M. Raissi, P. Perdikaris, and G. E. Karniadakis,
``Physics-informed neural networks: A deep learning framework
for solving forward and inverse problems involving nonlinear
partial differential equations,''
\textit{Journal of Computational Physics},
vol.~378, pp.~686--707, 2019.

\bibitem{Adler2018}
J. Adler and O. \"{O}ktem,
``Learned primal-dual reconstruction,''
\textit{IEEE Transactions on Medical Imaging},
vol.~37, no.~6, pp.~1322--1332, 2018.

\bibitem{Niemeyer2022}
M. Niemeyer, A. Barron, N. Mildenhall, M. Sajjadi,
A. Geiger, and A. Radwan,
``RegNeRF: Regularizing neural radiance fields for view synthesis
from sparse inputs,''
in \textit{Proc. IEEE/CVF Conf. Comput. Vis. Pattern Recognit. (CVPR)},
pp.~5480--5490, 2022.

\bibitem{Jain2021}
A. Jain, M. Tancik, and P. Abbeel,
``Putting NeRF on a diet: Semantically consistent few-shot
view synthesis,''
in \textit{Proc. IEEE/CVF Int. Conf. Comput. Vis. (ICCV)},
pp.~5885--5894, 2021.

\bibitem{schwarz2020graf}
K. Schwarz, A. Sauer, Y. Liao, and A. Geiger,
``GRAF: generative radiance fields for 3D-aware image synthesis,''
in \textit{Adv. Neural Inf. Process. Syst. (NeurIPS)}, 2020.

\bibitem{Chung2022}
H. Chung and J. C. Ye,
``Score-based diffusion models for accelerated MRI,''
\textit{Medical Image Analysis},
vol.~80, p.~102479, 2022.

\bibitem{kingma2014adam}
D. P. Kingma and J. Ba,
``Adam: a method for stochastic optimization,''
in \textit{Proc. Int. Conf. Learn. Represent. (ICLR)}, 2015.

\bibitem{otsu1979threshold}
N. Otsu,
``A threshold selection method from gray-level histograms,''
\textit{IEEE Trans. Syst. Man Cybern.}, vol. 9, no. 1,
pp. 62--66, 1979.

\bibitem{setio2017validation}
A. A. A. Setio \textit{et al.},
``Validation, comparison, and combination of algorithms for automatic detection of pulmonary nodules in computed tomography images,''
\textit{Med. Image Anal.}, vol. 42, pp. 1--13, 2017.

\bibitem{kundrapu2025kidney}
V. Kundrapu, T. R. Mettukuru, P. Narisetty, and T. Singh,
``Kidney Stone Detection in CT Scans: A Hybrid Approach with Machine Learning and Deep Learning,''
\textit{IFIP Adv. Inf. Commun. Technol.}, Springer Nature Switzerland, 2025,
doi: 10.1007/978-3-031-98360-3\_10.

\bibitem{srikanth2022xai}
K. S. Srikanth, T. K. Ramesh, S. Palaniswamy, and R. Srinivasan,
``XAI based model evaluation by applying domain knowledge,''
in \textit{Proc. IEEE Int. Conf. Electron. Comput. Commun. Technol. (CONECCT)}, IEEE, 2022.

\bibitem{afnaan2025brain}
K. Afnaan \textit{et al.},
``Deep learning for enhanced delineation and classification in brain MRI images,''
\textit{IFIP Adv. Inf. Commun. Technol.}, Springer Nature Switzerland, 2025,
doi: 10.1007/978-3-031-98356-6\_11.

\bibitem{afnaan2025gait}
K. Afnaan \textit{et al.},
``Leveraging convolutional neural networks for gait recognition and individual identification for improved neurological care,''
\textit{IFIP Adv. Inf. Commun. Technol.}, Springer Nature Switzerland, 2025,
doi: 10.1007/978-3-031-98356-6\_10.

\end{thebibliography}

\end{document}